\documentclass[fleqn,usenatbib]{mnras}

\usepackage{newtxtext,newtxmath}
\usepackage[T1]{fontenc}

\DeclareRobustCommand{\VAN}[3]{#2}
\let\VANthebibliography\thebibliography
\def\thebibliography{\DeclareRobustCommand{\VAN}[3]{##3}\VANthebibliography}

\usepackage{graphicx}	% Including figure files
\usepackage{amsmath}	% Advanced maths commands

\newcommand{\kms}{\mbox{$\mathrm{km\,s^{-1}}$}}
\newcommand{\MSUN}{\mbox{$\mathrm{M_{\odot}}$}}
\newcommand{\MJUP}{\mbox{$\mathrm{M_\mathrm{Jup}}$}}
\newcommand{\RSUN}{\mbox{$\mathrm{R_{\odot}}$}}

\title[Eclipsing white dwarf + brown dwarf binaries]{Two almost planetary mass survivors of common envelope evolution}

\author[S.~G.~Parsons et al.]{S.~G.~Parsons,$^{1}$\thanks{E-mail: s.g.parsons@sheffield.ac.uk}
A.~J.~Brown,$^{2}$
S.~L.~Casewell,$^{3}$
S.~P.~Littlefair,$^{1}$
J.~van Roestel,$^{4}$
A.~Rebassa-Mansergas,$^{5,6}$ \newauthor
R.~Murillo-Ojeda,$^{7}$
M.~A.~Hollands,$^{8}$
M.~Zorotovic,$^{9}$
N.~Castro Segura,$^{8}$
V.~S.~Dhillon,$^{1,10}$
M.~J.~Dyer,$^{1}$ \newauthor
J.~A.~Garbutt,$^{1}$ 
M.~J.~Green,$^{11}$
D.~Jarvis,$^{1}$ 
M.~R.~Kennedy,$^{12}$
P.~Kerry,$^{1}$
J.~McCormac,$^{8}$
J.~Munday,$^{8}$ \newauthor
I.~Pelisoli,$^{8}$
E.~Pike$^{1}$
and D.~I.~Sahman$^{1}$
\\
$^{1}$Astrophysics Research Cluster, School of Mathematical and Physical Sciences, University of Sheffield, Sheffield S3 7RH, UK\\
$^{2}$ Hamburger Sternwarte, University of Hamburg, Gojenbergsweg 112, 21029 Hamburg, Germany \\
$^{3}$School of Physics and Astronomy, University of Leicester, University Road, Leicester LE1 7RH, UK\\
$^{4}$Anton Pannekoek Institute for Astronomy, University of Amsterdam, 1090 GE, Amsterdam, The Netherlands\\
$^{5}$Departament de F{\'i}sica, Universitat Polit{\`e}cnica de Catalunya, c/Esteve Terrades 5, 08860 Castelldefels, Spain\\
$^{6}$Institut d'Estudis Espacials de Catalunya, c/ Esteve Terradas, 1, Edifici RDIT, Campus PMT-UPC, 08860 Castelldefels, Spain\\
$^{7}$Centro de Astrobiolog{\'i}a (CAB), CSIC-INTA, Camino Bajo del Castillo s/n, campus ESAC, 28692 Villanueva de la Ca{\~n}ada, Madrid, Spain. \\
$^{8}$Department of Physics, University of Warwick, Gibbet Hill Road, Coventry, CV4 7AL, UK\\
$^{9}$Instituto de F{\'i}sica y Astronom{\'i}a, Universidad de Valpara{\'i}so, Av. Gran Breta{\~n}a 1111, Valpara{\'i}so, Chile\\
$^{10}$Instituto de Astrofisica de Canarias, E38205 La Laguna, Tenerife, Spain\\
$^{11}$Max-Planck-Institut f{\"u}r Astronomie, K{\"o}nigstuhl 17, D-69117 Heidelberg, Germany\\
$^{12}$School of Physics, University College Cork, Cork, T12 K8AF, Ireland
}

\date{Accepted 2025 January 23. Received 2025 January 23; in original form 2024 December 17}

\pubyear{2025}

\begin{document}
\label{firstpage}
\pagerange{\pageref{firstpage}--\pageref{lastpage}}
\maketitle

\begin{abstract}
White dwarfs are often found in close binaries with stellar or even substellar companions. It is generally thought that these compact binaries form via common envelope evolution, triggered by the progenitor of the white dwarf expanding after it evolved off the main-sequence and engulfing its companion. To date, a handful of white dwarfs in compact binaries with substellar companions have been found, typically with masses greater than around 50\,{\MJUP}. Here we report the discovery of two eclipsing white dwarf plus brown dwarf binaries containing very low mass brown dwarfs. ZTF\,J1828+2308 consists of a hot ($15900\pm75$\,K) $0.610\pm0.004$\,{\MSUN} white dwarf in a 2.7 hour binary with a $0.0186\pm0.0008$\,{\MSUN} ($19.5\pm0.8$\,{\MJUP}) brown dwarf. ZTF\,J1230$-$2655 contains a cool ($10000\pm110$\,K) $0.65\pm0.02$\,{\MSUN} white dwarf in a 5.7 hour binary with a companion that has a mass of less than 0.0211\,{\MSUN} (22.1\,{\MJUP}). While the brown dwarf in ZTF\,J1828+2308 has a radius consistent with its mass and age, ZTF\,J1230$-$2655 contains a roughly 20 per cent overinflated brown dwarf for its age. We are only able to reconstruct the common envelope phase for either system if it occurred after the first thermal pulse, when the white dwarf progenitor had already lost a significant fraction of its original mass. This is true even for very high common envelope ejection efficiencies ($\alpha_\mathrm{CE}\sim 1$), unless both systems have extremely low metallicities. It may be that the lowest mass companions can only survive a common envelope phase if it occurs at this very late stage.
\end{abstract}


%%%%%%%%%%%%%%%%%%%%%%%%%%%%%%%%%%%%%%%%%%%%%%%%%%

%%%%%%%%%%%%%%%%% BODY OF PAPER %%%%%%%%%%%%%%%%%%

\section{Introduction}

Present day white dwarfs are the descendants of mostly A and F type stars, of which around 50 per cent are found in binary systems \citep[e.g.][]{DeRosa14}. It is therefore unsurprising that a significant fraction of white dwarfs are found in binaries \citep{Toonen17}. In many of these systems the separation between the white dwarf and its companion is smaller than the size of the white dwarf progenitor when it was a giant, leading to the idea that these binaries must have originally been wider and they were subsequently brought closer as a result of the evolution of the white dwarf progenitor. It is generally accepted that this orbital shrinkage is usually the result of a common envelope (CE) phase \citep{Paczynski76,Webbink84,Ivanova13}, caused by the giant star filling its Roche lobe and transferring mass to its companion. If the companion star cannot adjust quickly enough to this additional material then it too will fill its Roche lobe, resulting in an envelope of material surrounding the binary. This envelope will not necessarily be co-rotating with the binary and hence the stars experience a frictional force that extracts orbital energy from the system, causing the stars to spiral in towards each other. If sufficient orbital energy is transferred into the envelope then it can be ejected, leaving behind the now closer binary. If there is insufficient orbital energy, or energy is transferred from the orbit to the envelope very inefficiently, then the binary may instead merge during this phase.

Surviving the CE phase can be particularly challenging for substellar companions, such as brown dwarfs and planetary mass objects. Their lower masses mean they have less orbital energy available to eject the envelope. Nevertheless, there are a small number of close white dwarf plus brown dwarf binaries known (11 at present, see \citealt{French24} for an up to date list), with around 0.1-0.5 per cent of white dwarfs thought to have brown dwarf companions \citep{Farihi05,Steele11,Rebassa19}. This low number may reflect the difficulty in surviving the CE phase, but is primarily a result of the intrinsic rarity of brown dwarfs in binaries with main sequence stars with separations smaller than a few AU (i.e. the progenitors of close white dwarf plus brown dwarf binaries), the so-called "brown dwarf desert" \citep{Grether06,Grieves17}.

Brown dwarfs that do survive the CE phase and go on to closely orbit white dwarfs are extremely useful probes of binary evolution \citep{Zorotovic22}. When angular momentum loss through gravitational radiation brings these systems close enough for the brown dwarf to fill its Roche lobe, the system will form a cataclysmic variable (CV, \citealt{Warner95,Knigge11}). These binaries therefore directly create systems that resemble period-bounce CVs, the end state of CVs that began mass transfer with more massive donor stars.

Brown dwarfs in close (orbital periods of less than a few days) binaries with white dwarfs can also be highly irradiated by the white dwarf, which can have effective temperatures 10-100 times larger than the brown dwarf. Since the brown dwarfs are also tidally locked, there are significant temperature variations between the hemisphere facing towards the white dwarf (the `day' side) and the hemisphere facing away from white dwarf (the `night' side). Temperature differences of hundreds of Kelvin are possible \citep{Casewell15,Hallakoun23}, making these objects extreme analogues of hot Jupiters. Given the difficulty in directly characterising the atmospheres of hot Jupiters, which are significantly outshone by their host stars at all wavelengths, the atmospheres of brown dwarfs irradiated by white dwarfs can be directly probed at infrared wavelengths, where they can outshine the white dwarf \citep[e.g.][]{Casewell15,Longstaff17,Lew22,French24}, providing some insight into the otherwise-inaccessible atmospheric physics of hot Jupiter planets.

To date, the majority of close white dwarf plus brown dwarf binaries have been discovered by identifying apparently single white dwarfs with infrared excesses \citep[e.g.][]{Farihi05,Steele11,Girven11}. Given their small separations, these binaries are often found to be eclipsing, allowing precise measurements of the stellar and binary parameters \citep[e.g.][]{Littlefair14}. These studies typically find brown dwarf masses above around 50\,{\MJUP}, potentially indicating that this is the lower limit for the mass of an object that can survive a CE phase (i.e. avoid merging with the core of the giant star). However, the recent discovery of the transiting planet candidate around the white dwarf WD\,1856+534 in a 1.4 day period \citep{Vanderburg20} demonstrates that lower mass objects ($<$14\,{\MJUP} for WD\,1856+534b) can be found in close orbits around white dwarfs. While it is likely that this planet was scattered into its close orbit \citep{Vanderburg20}, it is also possible that it survived a CE phase \citep{Lagos21}, with the previous lower limit of around 50\,{\MJUP} being a selection effect caused by the requirement for a detectable infrared excess. Lower mass (hence fainter and cooler) brown dwarfs and planetary mass objects can be completely outshone by white dwarfs even at infrared wavelengths and may only be detected if they happen to eclipse the white dwarf (or possibly via high precision radial velocity monitoring of apparently single white dwarfs). 

The number of eclipsing white dwarf binaries has rapidly expanded in the last few years, thanks mainly to large scale, wide-field time-domain photometric sky surveys, such as the Zwicky Transient Facility (ZTF, \citealt{Bellm19,Graham19,Masci19}), with hundreds of eclipsing systems now known \citep[e.g.][]{VanRoestel19,Keller22,Kosakowski22,Brown23}. A small number of these eclipsing white dwarf systems sit on the white dwarf cooling track in the Gaia colour-magnitude diagram, implying that the companion to the white dwarf in these systems contributes a negligible amount of optical light compared to the white dwarf, making these excellent close white dwarf plus brown dwarf binary candidates.

In this paper we present follow up phase-resolved spectroscopy and high-speed photometry of two candidate white dwarf plus brown dwarf binaries. These were initially identified as white dwarfs with deep eclipses in ZTF and lacked any detectable excess flux at longer wavelengths, implying low mass or even substellar companions. Here we measure their stellar and binary parameters and find that both systems contain brown dwarfs with masses far below what is typically found in post-CE systems and so we test if standard CE reconstruction methods are able to reproduce these systems.

\section{Observations and their reduction}

\begin{table*}
 \centering
  \caption{Journal of observations.}
  \label{tab:obslog}
  \begin{tabular}{@{}lcccccc@{}}
  \hline
  Telescope/ & Date & Filters & No. of    & Exposure  & Transparency & seeing \\
  Instrument &      &         & exposures & times (s) &             & (arcsec) \\
  \hline
\multicolumn{3}{l}{\bf ZTF\,J1230$-$2655:} \\
  VLT/X-shooter & 2024-04-10 & UVB/VIS/NIR & 7/6/31   & 510/600/120 & Clear & 0.7--1.0 \\
  VLT/X-shooter & 2024-04-11 & UVB/VIS/NIR & 20/18/92 & 510/600/120 & Clear & 0.5--0.7 \\
  VLT/X-shooter & 2024-04-12 & UVB/VIS/NIR & 7/6/31   & 510/600/120 & Thin clouds & 1.0--1.2 \\  
  NTT/ULTRACAM  & 2024-05-02 & $u_s$,$g_s$,$i_s$ & 265/796/796 & 13.5/4.5/4.5 & Clear & 1.2--1.5 \\
\multicolumn{3}{l}{\bf ZTF\,J1828+2308:} \\
  VLT/X-shooter & 2024-04-10 & UVB/VIS/NIR & 10/9/46   & 510/600/120 & Clear & 0.7--1.0 \\
  VLT/X-shooter & 2024-04-11 & UVB/VIS/NIR & 10/9/46   & 510/600/120 & Clear & 0.5--0.7 \\
  GTC/HiPERCAM  & 2024-04-30 & $u_s$,$g_s$,$r_s$,$i_s$,$z_s$ & 565/1695/1695/847/847 & 6/2/2/4/4 & Clear & 0.7--1.0 \\
  \hline
\end{tabular}
\end{table*}

In this section we summarise our spectroscopic and photometric observations and their reduction. A full list of our observations is given in Table~\ref{tab:obslog}.

\subsection{X-Shooter spectroscopy}

We obtained phase-resolved medium resolution spectroscopy of our targets with the echelle spectrograph X-shooter \citep{Vernet11}, which is mounted at the Cassegrain focus of the VLT-UT2 at Paranal, Chile, on the nights of 2024 April 10-12. X-shooter covers the spectral range from the atmospheric cutoff in the UV to the near-infrared K band in three separate arms, known as the UVB (0.30$-$0.56 microns), VIS (0.56$-$1.01 microns) and NIR (1.01$-$2.40 microns). Separate slit widths can be set for each arm and our observations were performed with slit widths of 1.0, 0.9 and 0.9 arcsec in the UVB, VIS and NIR arms respectively. We also binned the detector in the VIS arm by a factor of 2 in the spatial direction, while the UVB arm was binned by a factor of 2 in both the spatial and dispersion directions. This results in a resolution of R$\simeq$5000 in the UVB and R$\simeq$8000 in the VIS arm. In order to minimise slit losses we ran the active flexure compensation procedure once every hour and realigned the slit to the parallactic angle while doing so.

All of the data were reduced using the standard X-shooter pipeline release (version 3.6.3) within {\sc esoreflex} \citep{Freudling13}. Due to the faintness of our targets in the NIR arm (both are too faint at these wavelengths to be detected in the 2MASS survey) and the fact that we did not nod the telescope between observations, the data for both of our targets in the NIR are extremely poor and we do not include these data in any of our analysis.

The standard X-shooter pipeline can reach a velocity accuracy of around 8\,{\kms} in the VIS arm. We improved this accuracy by fitting a number of telluric absorption features and corrected for small (typically $\sim$1\,{\kms}) systemic velocity offsets in the data via the method described in \citet{Parsons17}. This allowed us to reach an accuracy of a few {\kms} around the H$\alpha$ line. These corrections predominantly affect our measurement of the systemic velocity of our systems, since accurate wavelength calibration is essential for this. Differences in the accuracy of the wavelength calibration from spectrum to spectrum are minimal, meaning that our radial velocity semi-amplitude measurements can be more precise than this ($<$1\,{\kms}). All spectra were then placed on a barycentric wavelength scale.

\subsection{ULTRACAM and HiPERCAM photometry}

ZTF\,J123016.59$-$265551.34 (listed as GALEX\,J123016.6$-$265551 in SIMBAD, hereafter ZTF\,J1230$-$2655) was observed with the high-speed camera ULTRACAM \citep{Dhillon07}, mounted on the 3.5-m New Technology Telescope at La Silla, Chile on the night of 2024 May 2. ULTRACAM uses a triple beam setup and three frame transfer CCD cameras, which allows simultaneous data in three different wavebands with negligible (24\,ms) dead-time between exposures. For our observations we made use of the high throughput super-SDSS $u_s$, $g_s$ and $i_s$ filters and targeted the orbital phases around the eclipse of the white dwarf. We used exposure times of 4.5 seconds in the $g_s$ and $r_s$ bands and longer exposure times of 13.5 in the $u_s$ band.

We observed ZTF\,J182848.77+230838.0 (hereafter ZTF\,J1828+2308) with the high-speed quintuple band camera HiPERCAM \citep{Dhillon21} mounted on the 10.4-m Gran Telescopio Canar{\'i}as (GTC) on La Palma, Spain on 2024 April 30. HiPERCAM allows simultaneous observations in the super-SDSS $u_s$,$g_s$,$r_s$,$i_s$ and $z_s$ bands with negligible (8\,ms) dead-time between exposures. We used exposure times of 2 seconds in the $g_s$ and $r_s$ bands, 4 seconds in the $i_s$ and $z_s$ bands and 6 seconds in the $u_s$ band.

Both the ULTRACAM and HiPERCAM data were reduced using the HiPERCAM pipeline software\footnote{\url{https://github.com/HiPERCAM/hipercam}}. After bias subtraction and flat fielding (using twilight flats) the source flux was determined with aperture photometry using a variable aperture scaled according to the full width at half maximum (FWHM). The HiPERCAM $z_s$ band data were also defringed using a reference fringe map. Transmission variations were accounted for by determining the flux relative to a comparison star in the field of view. The data were then flux calibrated using observations of the standard stars WD\,1031$-$114 (for the ULTRACAM observations of ZTF\,J1230$-$2655) and Feige\,66 (for the HiPERCAM observations of ZTF\,J1828+2308), using the reference magnitudes for the super-SDSS filters from \citet{Brown22}. 

All times were converted to a TDB (Barycentric Dynamical Time) timescale corrected to the barycentre of the solar system. Using modified Julian dates format, this results in timing measurements of the form BMJD(TDB).

As noted by \citet{Brown23}, there is a faint source 2.8{\arcsec} away from ZTF\,J1828+2308. Fortunately due to the good seeing during our HiPERCAM observations ($\sim$0.8\arcsec) the target and this source are well resolved and we masked out any pixels around this source to ensure that the final light curve and flux calibration were not affected by it.

\section{Results}

\subsection{ZTF J182848.77+230838.0}

ZTF\,J1828+2308 was first listed as a candidate white dwarf plus brown dwarf binary by \citet{Kosakowski22}, who presented the ZTF light curve and noted the very deep eclipse of the white dwarf every 2.7 hours, as well as the lack of any clear excess in the PanSTARRS $y$ band, indicating a likely substellar companion. Follow up high-speed multi-band ULTRACAM photometry of this object was presented by \citet{Brown23}, who used various theoretical relationships to constrain the mass of the white dwarf companion from the light curves, concluding that it was very likely substellar.

Our HiPERCAM light curves achieve a much greater depth than the ULTRACAM data in \citet{Brown23}, but still fail to detect the companion in any bands. By combining all observations taken during the eclipse, we place limits on the companion's brightness in the redder bands (where we might expect to detect a brown dwarf) of $i_s > 24.5$ and $z_s > 24.1$. Based on the Gaia DR3 parallax ($4.91\pm0.10$\,mas, \citealt{Gaia23}), this corresponds to absolute magnitude limits of $M_i > 17.9$ and $M_z > 17.6$, implying a spectral type later than L4 \citep{Kiman19}.

While we do not detect the underlying photosphere of the companion to the white dwarf, we do detect H$\alpha$ emission from its heated face. This, along with the radial velocity variations of the white dwarf itself, allow us to place strong constraints on the stellar masses, which combined with analysis of the white dwarf spectrum and eclipse light curves yield the full system parameters.

\subsubsection{Radial velocities}

\begin{figure*}
  \begin{center}
    \includegraphics[width=\textwidth]{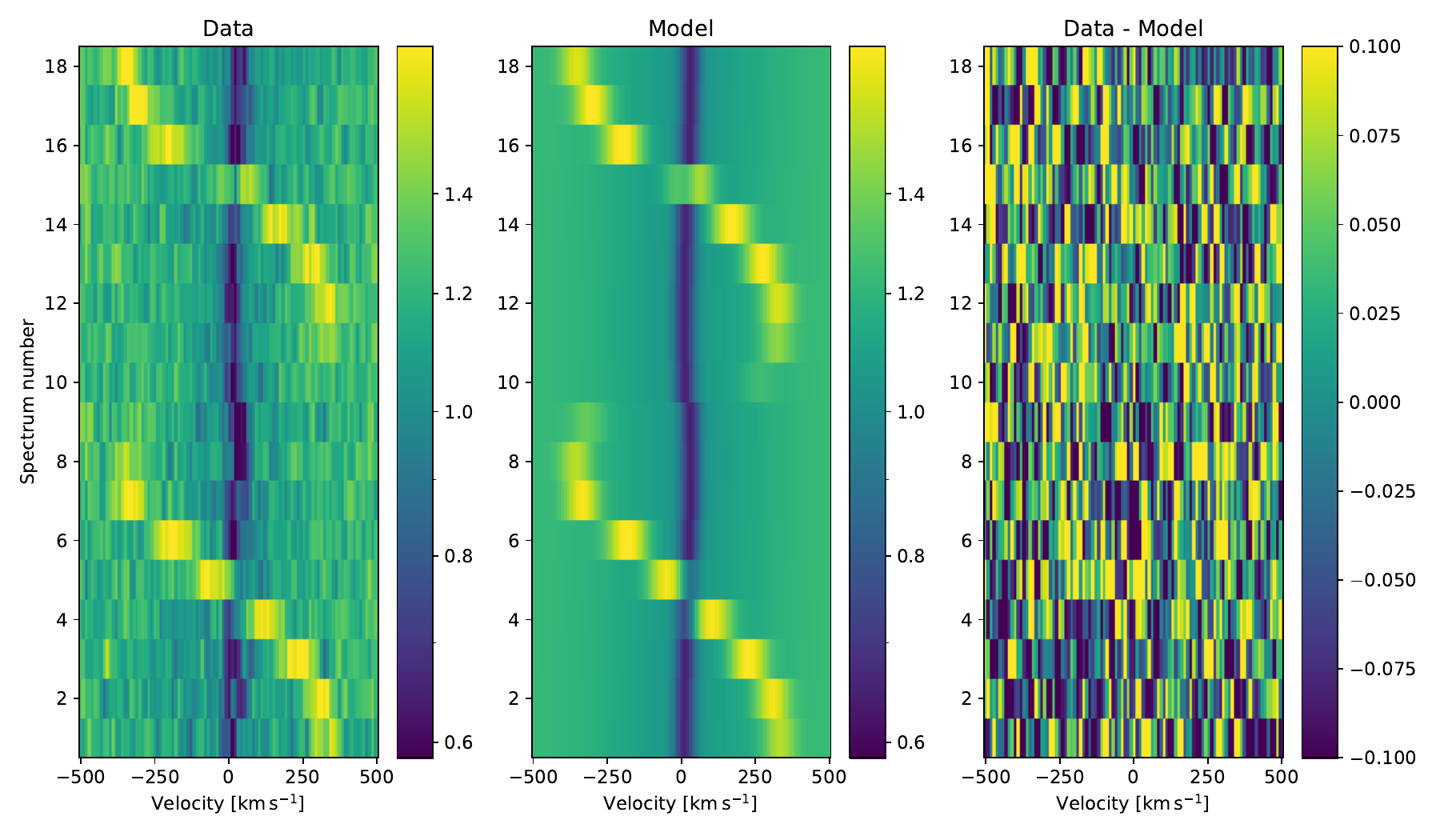}
    \caption{Trailed spectra of the H$\alpha$ line in ZTF\,J1828+2308. {\it Left:} the original data, where the non-LTE core of the white dwarf absorption line is clear as well as emission from the companion moving in anti-phase. The slight velocity jump between spectrum 9 and 10 is due to the fact that the data were taken on different nights and some orbital phases are missed. {\it Centre:} our best fit model to the data. {\it Right:} the residuals to the fit.}
  \label{fig:1828_trail}
  \end{center}
\end{figure*}

ZTF\,J1828+2308 contains a fairly hot ($T_\mathrm{eff}=15900$\,K) white dwarf. This, combined with the short orbital period of 2.7 hours (hence small orbital separation), means that the companion is moderately irradiated. The effect is fairly minor and there is not a clear reflection effect in the ZTF light curves or follow up high-speed data (although these are focused around the white dwarf eclipse phases, where irradiation effects are minimal). However, inspection of the X-shooter spectra clearly revealed H$\alpha$ emission moving at high velocities and in anti-phase with the white dwarf absorption line (see Figure~\ref{fig:1828_trail}). The emission is strongest around phase 0.5 (roughly spectra 5 and 15 in Figure~\ref{fig:1828_trail}) and weakens towards phase 1 (the eclipse of the white dwarf). Note that there is a small gap in the coverage of our X-shooter data between orbital phases 0.8 and 1.15 (i.e. around the eclipse of the white dwarf), hence the slight jump in the data between spectrum 9 and 10 in Figure~\ref{fig:1828_trail}.

While subtle, we also clearly detect the motion of the white dwarf in the core of the H$\alpha$ line. Since the higher order Balmer lines lack this strong core, the H$\alpha$ line is the only feature that we could use to measure the white dwarf's velocity. We therefore focused on measuring the velocities of both stars by fitting the H$\alpha$ line in all our spectra.

We fitted the H$\alpha$ line following the method of \citet{Parsons17b}, in which the line is fitted with a combination of a first order polynomial and three Gaussian components. Two Gaussian components with the same velocity are used for the white dwarf absorption line, a wide component for the wings of the line and a narrower component for the core. The third Gaussian component is used for the emission from the companion star. All of the spectra are fitted simultaneously, with the white dwarf absorption components changing position according to $\gamma_\mathrm{WD}$ + K$_\mathrm{WD} \sin{(2 \pi \phi)}$, where $\phi$ is the orbital phase (note that this results in K$_\mathrm{WD}$ having a negative value due to setting $\phi=0$ as the mid-eclipse of the white dwarf, but only the amplitude is important, so we drop the sign when reporting this value), and the emission component from the companion changing position according to $\gamma_\mathrm{BD}$ + K$_\mathrm{emis} \sin{(2 \pi \phi)}$. The widths and heights of the two white dwarf components remain the same in all spectra, while the height of the emission component from the companion is modulated as $(1 - \cos{\phi})/2$, to simulate the irradiation effect (its width remains the same in all spectra).

We used the ephemeris from \citet{Brown23} to determine the orbital phase of each spectrum, $\phi = (T_\mathrm{obs}-T_0)/P_\mathrm{orb}$, where $T_\mathrm{obs}$ is the mid-exposure time of the spectrum, $T_0$ is the time of mid-eclipse of the white dwarf and $P_\mathrm{orb}$ is the orbital period, but also allowed the values to vary within the quoted uncertainties by placing Gaussian priors on $T_0$ and $P_\mathrm{orb}$. The fit was performed using the Markov chain Monte Carlo (MCMC) method as implemented in the {\sc emcee} Python package \citep{Foreman13}. We used 100 walkers, with a burn-in period of 1500, and 5000 production steps.

The best fit model and residuals to the fit are shown in the central and right-hand panels of Figure~\ref{fig:1828_trail}. We found that the radial velocity semi-amplitude of the white dwarf (K$_\mathrm{WD}$) is extremely small ($10.5\pm1.1$\,\kms) despite the short period and high inclination of the binary, strongly implying a very low mass for the companion. This is also consistent with the large emission-line velocity for the companion star ($343.4\pm1.2$\,\kms), although since this emission originates from the inner hemisphere of the companion it does not track the centre of mass of the star, but the centre of light of the emission region. As such it represents a lower limit on the true radial velocity of the companion ($K_\mathrm{BD}$), which are related via
\begin{equation}
K_\mathrm{BD} = \frac{K_\mathrm{emis}}{1 - f(1+q) R_\mathrm{BD}/a}, \label{eqn:kcorr}
\end{equation}
where $q=M_\mathrm{BD}/M_\mathrm{WD}$ is the mass ratio, $R_\mathrm{BD}/a$ is the radius of the companion star scaled to the semi-major axis and $f$ is is a constant between 0 and 1 which depends upon the location of the centre of light \citep{Wade88,Parsons12}. In the most extreme case ($f=1$) the emission is assumed to entirely originate at the point on the companion star's surface closest to the white dwarf (the substellar point), while $f=0$ assumes that the emission is uniform over the surface of the star and hence does track the centre of mass of the star. Using $f=0$ (i.e. $K_\mathrm{BD} = K_\mathrm{emis}$) gives an upper limit to the mass ratio of $q < K_\mathrm{WD} / K_\mathrm{emis} = 0.03$, significantly smaller than what is typically found in white dwarf plus brown dwarf binaries ($q\sim0.1$), confirming the very low mass of the companion star (alternatively, a very high white dwarf mass is possible, although we will see in the next sections that this is not the case).

\subsubsection{White dwarf spectral fit}

\begin{figure}
  \begin{center}
    \includegraphics[width=\columnwidth]{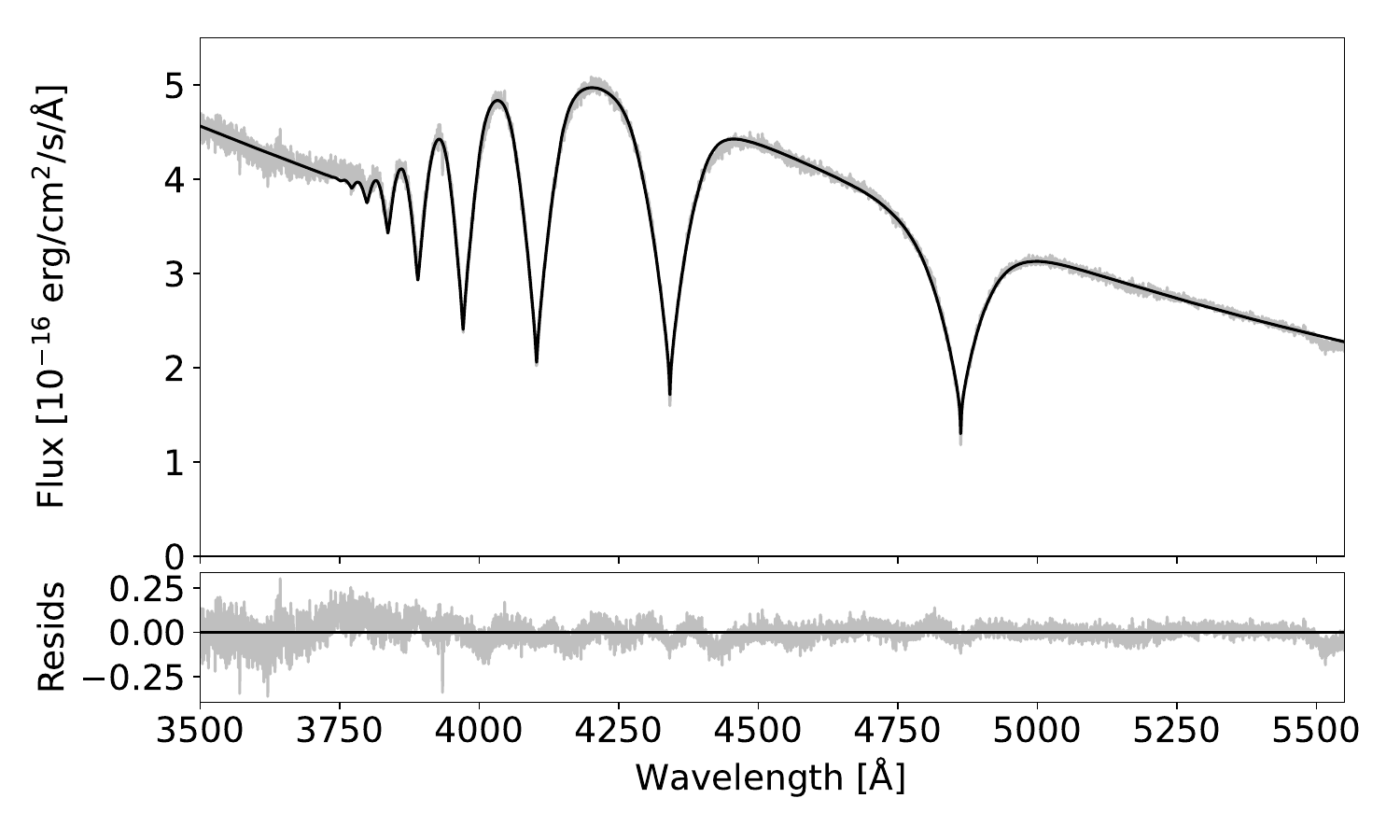}
    \caption{Averaged X-shooter UVB arm spectrum of ZTF\,J1828+2308, where all spectra have been shifted into the white dwarf frame before averaging. The best fit \citet{Koester10} model is overplotted in black and the residuals to the fit are shown in the lower panel. Note that the only spectroscopic feature detected from the companion to the white dwarf is weak H$\alpha$ emission, no other hydrogen lines show detectable emission components.}
  \label{fig:1828_specfit}
  \end{center}
\end{figure}

Having calculated the white dwarf's radial velocity semi-amplitude, we shifted all our spectra into the rest frame of the white dwarf and averaged them together to produce a high quality spectrum of the white dwarf. We then fitted this white dwarf spectrum using \citet{Koester10} models in order to constrain the effective temperature and surface gravity. We also included a polynomial curve in the fit to account for slit losses and the imperfect flux calibration of the X-shooter data. The models were convolved with a Gaussian to match the resolution of the data before fitting. Our short exposure times (and the small velocity of the white dwarf) means that the movement of the white dwarf during each exposure is negligible and hence there is no additional smearing of the spectral features. We do not include any corrections for interstellar reddening.

The average UVB arm spectrum and best fit model for ZTF\,J1828+2308 are shown in Figure~\ref{fig:1828_specfit}. We find $T_\mathrm{eff}=15900\pm75$\,K and $\log{(g)}=8.05\pm0.05$, where systematic uncertainties have been applied to both values based on \citet{Hollands24}. We found that applying the 3D corrections of \citet{Tremblay13} made no difference to the parameters of this white dwarf due to its temperature being too high for convection. Assuming the mass-radius relationship from \citet{Bedard20}, this implies a white dwarf mass of around 0.6\,{\MSUN}. These values are consistent with those found by \citet{Brown23} from analysis of multi-band eclipse light curves.

We also detected a weak Ca\,{\sc ii} 3934\,{\AA} absorption line in the spectra of ZTF\,J1828+2308, but do not detect any velocity variations from this line. It is also very narrow leading us to conclude that this is most likely an interstellar absorption feature. No other lines except the hydrogen Balmer series are detected in the X-shooter spectrum of ZTF\,J1828+2308.

\subsubsection{Light curve fit}

\begin{figure}
  \begin{center}
    \includegraphics[width=\columnwidth]{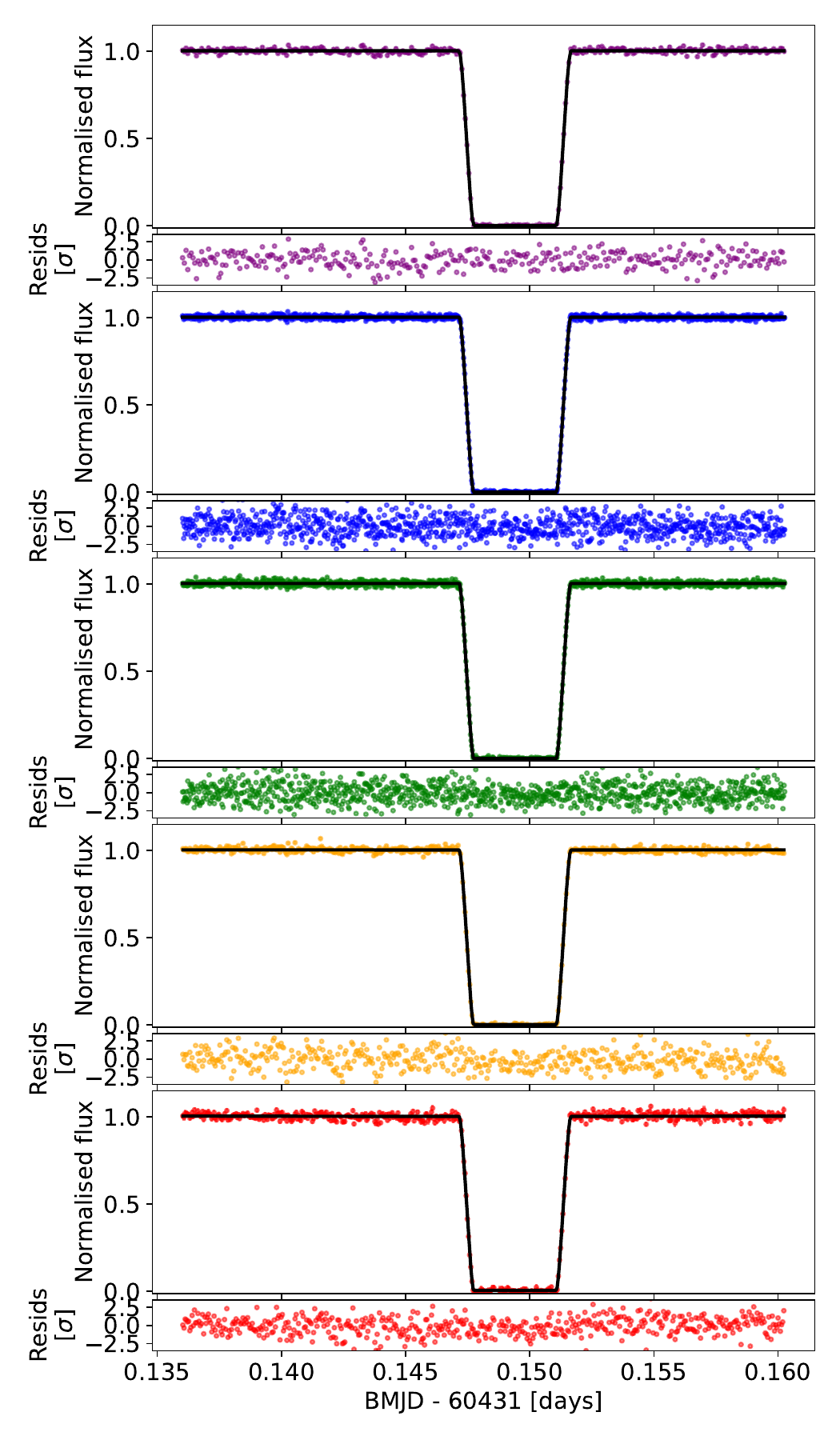}
    \caption{GTC+HiPERCAM eclipse light curves of ZTF\,J1828+2308 (top to bottom: $u_s$, $g_s$, $r_s$, $i_s$ and $z_s$ bands) along with the best fit models (black lines). Residuals to the fits are shown beneath each panel (in standard deviations).}
  \label{fig:1828_lcurve}
  \end{center}
\end{figure}

Our high-speed, multi-band HiPERCAM light curves of the eclipse of the white dwarf in ZTF\,J1828+2308 show an extremely deep eclipse in all bands (no flux is detected from the companion during the eclipse in any band, see Figure~\ref{fig:1828_lcurve}). The shape of the eclipse is sensitive to the sizes of the stars as well as the inclination of the binary. However, in general there is not sufficient information in the eclipse light curve alone to simultaneously solve for the radii of both stars and also the orbital inclination \citep{Parsons17}. This can be overcome by also including any additional constraints from the spectroscopic analysis when fitting the light curves. Of particular importance for this system is the measurement of the surface gravity of the white dwarf from the spectral fit and the constraints on the radial velocities of the two stars. This is because knowledge of both radial velocities yields the masses for any given inclination ($i$) via Kepler's third law,
\begin{equation}
\frac{P_\mathrm{orb} K_{1,2}^3}{2 \pi G} = \frac{M_{2,1} \sin^3{i}}{(1+q)^2,} \label{eqn:kep3}
\end{equation}
where the subscripts 1 and 2 refer to the two stars (i.e. $K_1$ is used with $M_2$ and vice versa) and $q=K_1/K_2$. At the same time, the shape of the eclipse yields the radii of the two stars for a given inclination, so combining these two pieces of information means that we can determine the surface gravity of the white dwarf at any inclination. We can then use the spectroscopically determined value of $\log{g}$ to determine the correct inclination and hence stellar and binary parameters.

Unfortunately we do not have a direct measurement of the radial velocity of the companion to the white dwarf, but a lower limit from the H$\alpha$ emission line. We can use Equation~\ref{eqn:kcorr} to correct the emission line velocity to the centre of mass velocity, but a value of $f$ must be assumed (for a given inclination $R_\mathrm{BD}/a$ is known from the eclipse shape and since $q=K_\mathrm{WD}/K_\mathrm{BD}$ it is possible to solve for this analytically). Therefore, in order to fit the eclipse and determine the stellar and binary parameters a value of $f$ must be chosen. Typically a value around 0.5 is used \citep[e.g.][]{Parsons17} which assumes that the emission is mostly uniformly spread over the heated face of the star. This is also what is measured in systems where there are both irradiation-driven emission lines and absorption lines from the secondary that track the centre of mass \citep{Parsons12b}. These studies find that different emission lines need different correction factors, based on the optical depth of the emission, with H$\alpha$ emission typically needing $f=0.5$ \citep{Parsons10}, although see \citet{vanRoestel17} for an alternative approach to determining this value. Therefore, we adopt a value of $f=0.5$ when fitting the light curves of ZTF\,J1828+2308 (although see the end of this section for an alternative approach).

With a value chosen for the $f$ term in Equation~\ref{eqn:kcorr} we were then able to fit the eclipse light curves to determine the stellar and binary parameters. The fit was performed using {\sc lcurve} \citep{Copperwheat10}, a code specifically designed for fitting the light curves of white dwarf binaries, which accounts for Roche distortion and irradiation. We fitted for the two masses, $M_\mathrm{WD}$ and $M_\mathrm{BD}$, the two radii, $R_\mathrm{WD}$ and $R_\mathrm{BD}$, the effective temperature of the white dwarf, $T_\mathrm{eff,WD}$, the orbital inclination, $i$, and the time of mid-eclipse, $T_0$. We held the orbital period fixed at the value from \citet{Brown23}.

{\sc lcurve} requires as input: the mass ratio, the orbital inclination, the radii of both stars scaled by the separation (the separation can be calculated from the masses and orbital period), the temperatures of the two stars (we fix the temperature of the companion to 1500\,K, but since it is undetected in the light curves this has no effect on the fit), the limb-darkening coefficients for both stars, which we set for the white dwarf by calculating its $\log{g}$ value from the mass and radius and interpolating the 4-parameter non-linear limb darkening values from \citet{Claret20} in each band. The limb darkening coefficients of the companion star have no effect on the fit, so we set these to a simple linear coefficient of 0.5 in all bands. {\sc lcurve} also requires a central eclipse time and an orbital period but all other parameters are irrelevant for this fit.

\begin{figure}
  \begin{center}
    \includegraphics[width=\columnwidth]{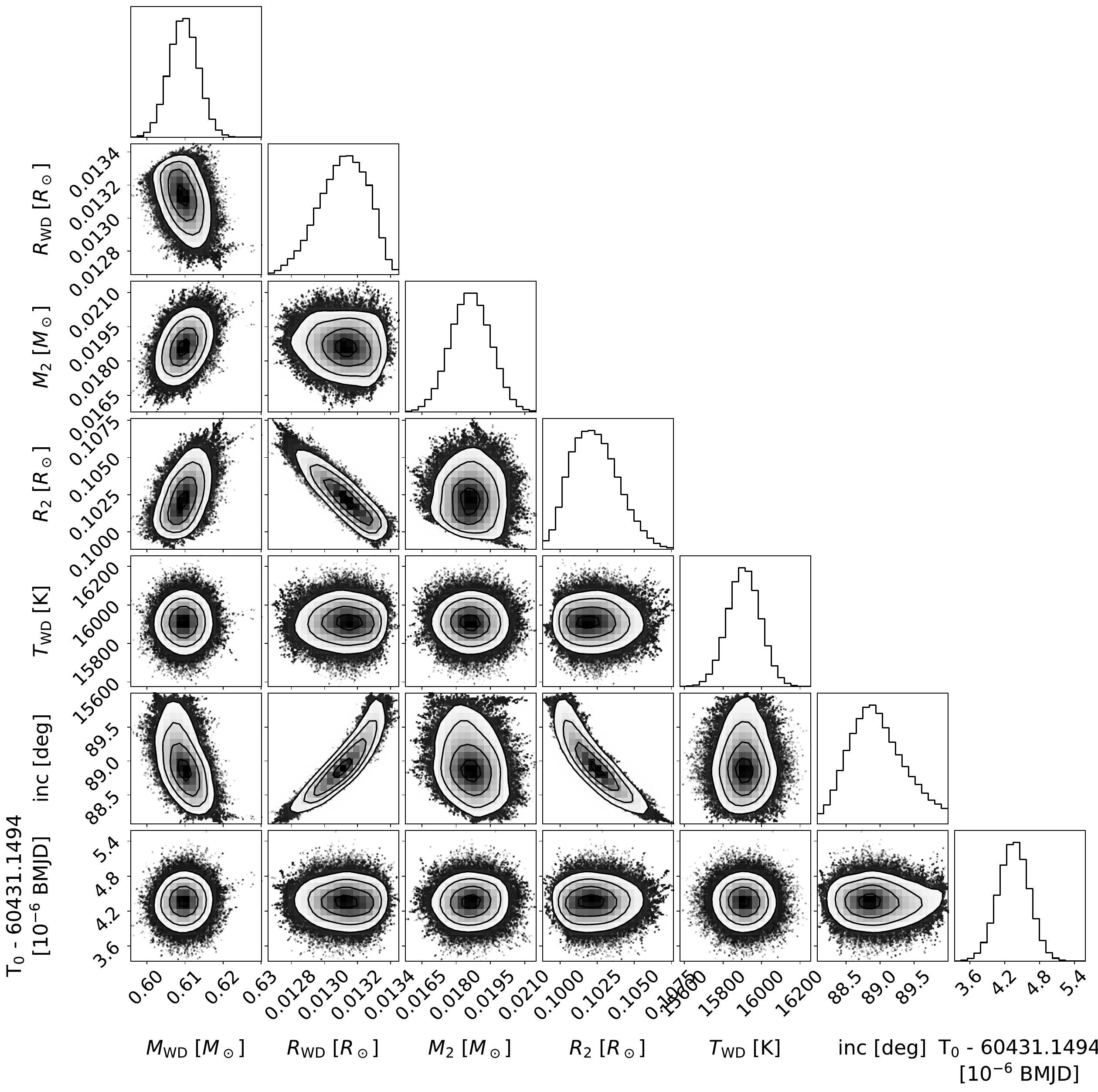}
    \caption{Posterior probability distributions for model parameters obtained through fitting the GTC+HiPERCAM light curves of ZTF\,J1828+2308.}
  \label{fig:1828_lcurve_corner}
  \end{center}
\end{figure} 

The light curves are only weakly dependent upon the temperature of the white dwarf. In principle the out-of-eclipse flux levels in each band constrain the temperature, but the accuracy of this is dependent on the quality of the flux calibration. Moreover, {\sc lcurve} computes models assuming black body temperatures, so a conversion must be made \citep[e.g.][]{Brown22}. Given that we already had a precise measurement for the white dwarf temperature from fitting the spectrum, we decided to normalise each of our light curves before fitting and constrain the white dwarf temperature via a Gaussian prior on the fit based on the spectroscopic value. This means that a conversion to a black body temperature is not required and also has the advantage that the temperature can still vary during the fit within the uncertainties from the spectroscopic fit, allowing a full exploration of the range of limb darkening coefficients for the white dwarf (which depend on its temperature). Given the extremely well sampled and high signal-to-noise ratio of our HiPERCAM light curves, the fit is sensitive to the limb darkening of the white dwarf and so we did not want to fix these values. The consequence of this is that the white dwarf temperature value (and uncertainty) returned by the fit is essentially just the spectroscopic value and so we do not improve on this with the light curve data. 

Along with the prior on the white dwarf temperature, we also placed a Gaussian prior on the surface gravity of the white dwarf, based on the spectroscopically measured value. Finally, for each model we also computed the expected values of $K_\mathrm{WD}$ and $K_\mathrm{BD}$ using Equation~\ref{eqn:kep3} and hence $K_\mathrm{emis}$ (based on Equation~\ref{eqn:kcorr} with $f=0.5$). We placed Gaussian priors on $K_\mathrm{WD}$ and $K_\mathrm{emis}$ based on the measured values and their uncertainties. We then performed an MCMC fit of the light curves in all band simultaneously using {\sc emcee}. We used 100 walkers, with a burn-in period of 2000 and 10000 production steps.

The HiPERCAM eclipse light curves along with the best fit model are shown in Figure~\ref{fig:1828_lcurve}, the best fit parameters are listed in Table~\ref{tab:params} and the MCMC parameter distributions are shown in Figure~\ref{fig:1828_lcurve_corner}. We also calculated the white dwarf cooling age from the \citet{Bedard20} models and the volume-averaged Roche lobe filling factor of the brown dwarf, both of which are listed in Table~\ref{tab:params}. We find that the companion to the white dwarf in ZTF\,J1828+2308 is a $0.0186\pm0.0008$\,{\MSUN} ($19.5\pm0.8$\,\MJUP) object. This makes the companion the second lowest mass object transiting a white dwarf after the planet candidate around WD\,1856+534 (excluding the transiting debris systems), although with a mass high enough that it is likely a brown dwarf rather than a planet.

\begin{figure}
  \begin{center}
    \includegraphics[width=\columnwidth]{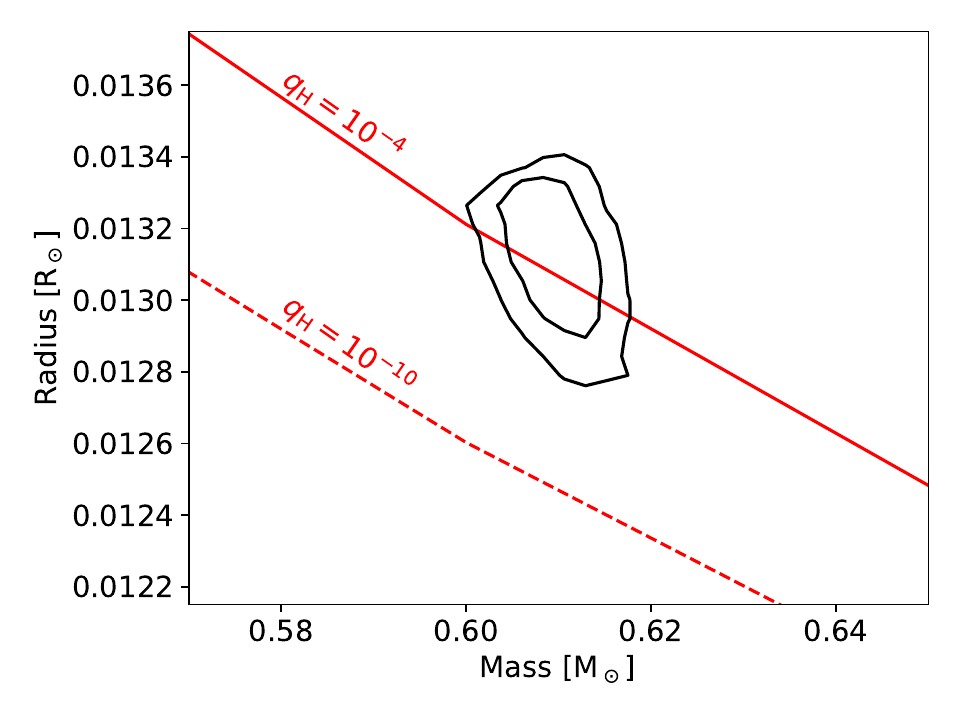}
    \caption{Constraints on the mass and radius of the white dwarf in ZTF\,J1828+2308 shown as contours (68 and 95 percentile regions). Also shown are theoretical mass-radius relationships for CO core white dwarfs from \citet{Bedard20} with thick hydrogen envelopes ($q_\mathrm{H}=10^{-4}$, solid line) and with thin hydrogen envelopes ($q_\mathrm{H}=10^{-10}$, dashed line). A very good agreement is found with the thick envelope models.}
  \label{fig:1828_massrad}
  \end{center}
\end{figure} 

We also find that the white dwarf has a mass of $0.610\pm0.004$\,{\MSUN} and a radius of $0.0131\pm0.0002$\,{\RSUN} and is therefore likely a CO core white dwarf (Figure~\ref{fig:1828_massrad}) that reached the asymptotic giant branch (AGB), in contrast to many known close white dwarf plus brown dwarf binaries that contain low-mass He core white dwarfs (see the list in \citealt{Zorotovic22} for examples). Figure~\ref{fig:1828_massrad} shows our mass and radius constraints relative to theoretical models for CO core white dwarfs with both thick and thin hydrogen envelopes \citep{Bedard20}. Our measurements are in excellent agreement with the thick envelope models, as is generally found for white dwarfs in close binaries \citep{Parsons17}.

As a final step, we tested the effects of our choice of $f=0.5$ as the emission-line correction factor. Due to the low values of the mass ratio ($q=0.03$) and the scaled radius of the companion ($R_\mathrm{BD}/a=0.12$) the choice of $f$ has only a minor effect on the final results (see Equation~\ref{eqn:kcorr}). This is because the companion is far from the centre of mass of the binary and small compared to the separation of the stars, so there is little difference in the velocity of its substellar point compared to its centre of mass. However, in order to see whether using the exact value of $f=0.5$ has any effect we repeated our fit of the eclipse light curves, but this time forced the white dwarf to follow the mass-radius relationship of \citet{Bedard20} but allowed $f$ to vary freely (between 0 and 1). In this case we found a best fit value of $f=0.52\pm0.02$ and all other parameters consistent with the previous fit. A value close to 0.5 is unsurprising, given the excellent agreement in our original fit between the white dwarf parameters and theoretical models in Figure~\ref{fig:1828_massrad}. We are therefore confident that the values for the physical parameters of the brown dwarf are robust against the choice of $f$.

\subsection{ZTF\,J123016.59$-$265551.34}

\begin{figure*}
  \begin{center}
    \includegraphics[width=0.45\textwidth]{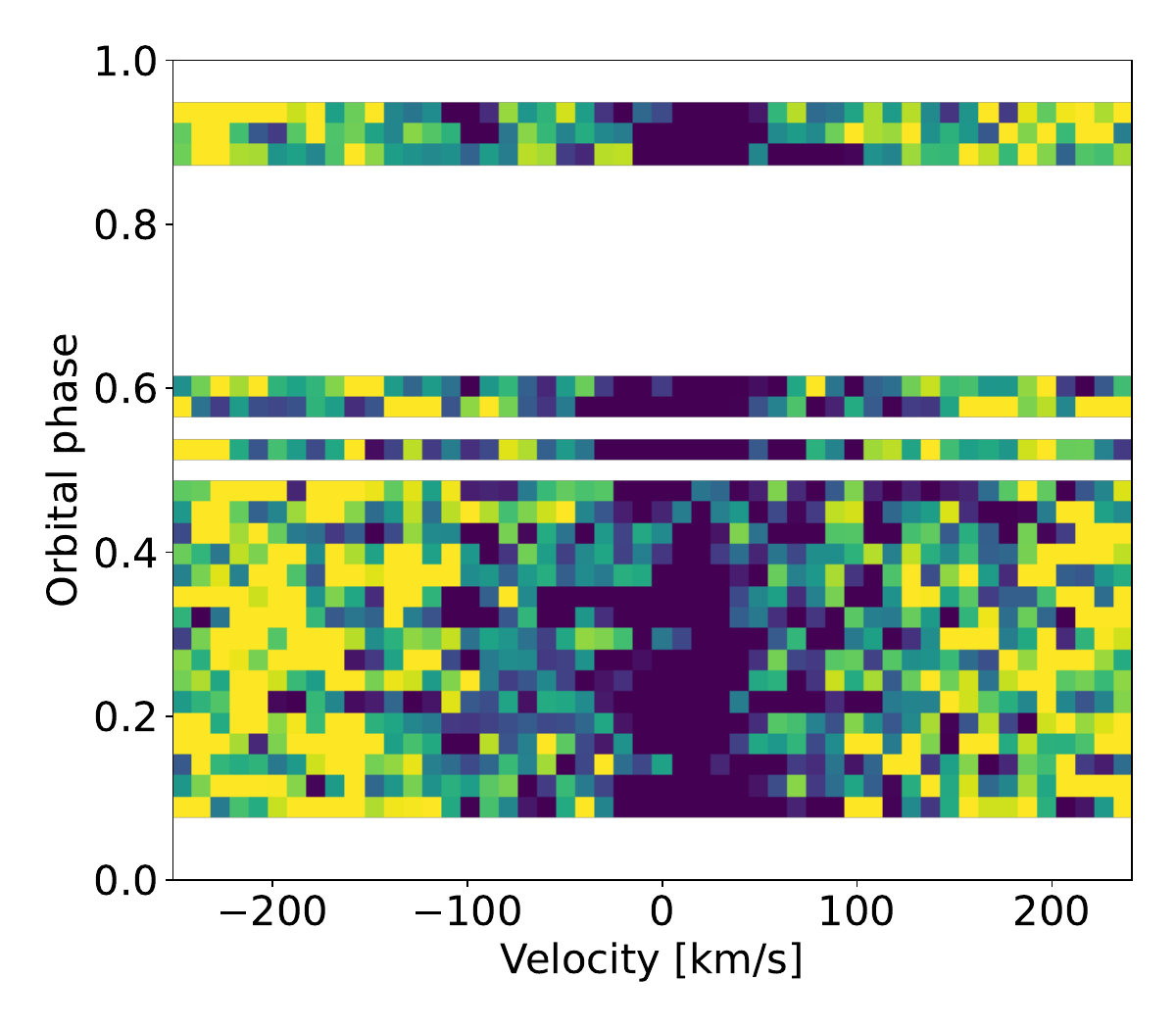}
    \includegraphics[width=0.545\textwidth]{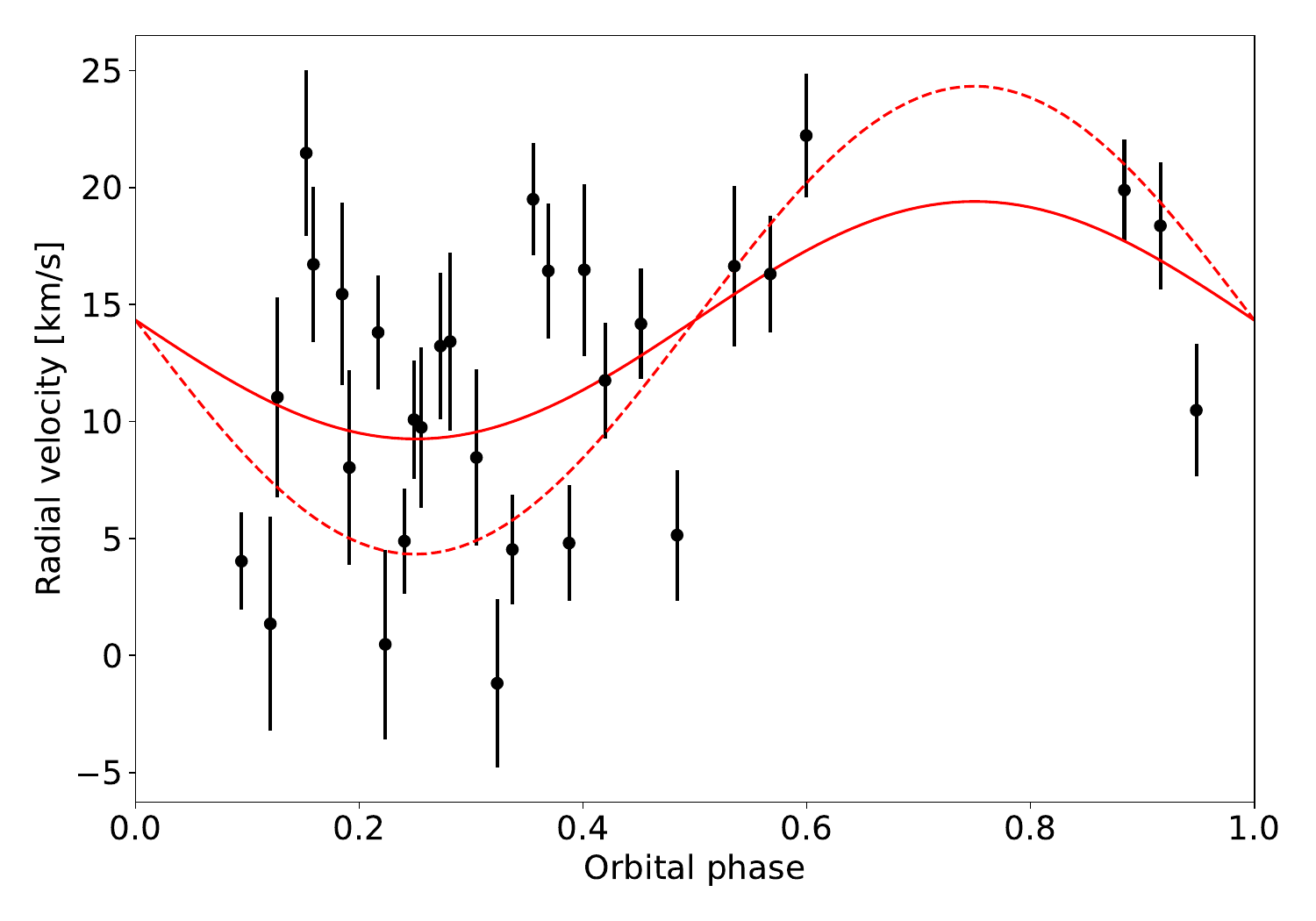}
    \caption{{\it Left:} phase-folded trailed spectra of the H$\alpha$ line in ZTF\,J1230$-$2655. The non-LTE core of the white dwarf absorption line is clear, but there is no obvious velocity variations. {\it Right:} radial velocity measurements from the H$\alpha$ absorption line along with a best fit (solid-red line), which is not statistically significant. Accounting for systematic uncertainties in these measurements of 5\,{\kms} (not included in the plotted errorbars) we determine an upper limit on the radial velocity amplitude of 9\,{\kms}, shown as the dashed-red line.}
  \label{fig:1230_trail}
  \end{center}
\end{figure*}

ZTF\,J1230$-$2655 was discovered as an eclipsing white dwarf binary using data from ZTF (van Roestel et~al. in prep). However, given its very southerly declination, the ZTF data coverage is fairly sparse and we made use of data from the Asteroid Terrestrial-impact Last Alert System (ATLAS, \citealt{Tonry18}) to improve the ephemeris measurements and confirm the eclipsing nature of the system. Given that ZTF\,J1230$-$2655 sits on the white dwarf cooling track in the Gaia DR3 colour-magnitude diagram the companion cannot contribute a significant amount of the optical light.

The companion to the white dwarf in ZTF\,J1230$-$2655 is undetected in our follow up ULTRACAM light curves and we place a limit of $i_s > 23.2$ based on all of our in-eclipse observations. Using the Gaia DR3 parallax ($5.11\pm0.25$\,mas, \citealt{Gaia23}) this corresponds to $M_i>16.7$, implying a spectral type of later than L2 \citep{Kiman19}.

\subsubsection{Radial velocities}

Our X-shooter data of ZTF\,J1230$-$2655 do not reveal any clear evidence for irradiation-induced emission lines from the companion star at H$\alpha$. Moreover, while the narrow absorption core of the H$\alpha$ line from the white dwarf is clear (see the left-hand panel of Figure~\ref{fig:1230_trail}), there are no obvious velocity variations. We attempted to use the same fitting procedure that we used for the spectra of ZTF\,J1828+2308 (removing the emission component), but failed to measure any significant velocity variations when fitting all the data simultaneously. We therefore opted instead to fit each spectrum individually with a combination of a first order polynomial and two Gaussian components (a wide component for the wings and a narrower one for the line core), forcing the two Gaussian components to have the same velocity.

The velocity measurements for the white dwarf in ZTF\,J1230$-$2655 are shown in the right-hand panel of Figure~\ref{fig:1230_trail}. The best fit sinusoid is also plotted, but the fit is poor and the data appear to just randomly scatter, rather than vary sinusoidally (although we unfortunately did not cover orbital phases near 0.75). In order to see if any sinusoidal signal is present in these data we performed two fits to the velocities, one in which we fitted a sinusoid to the data and another where we fitted the velocities as a constant value. In each case we also included an additional systematic error term in the fit and determined the model likelihood via
\begin{equation}
\ln \mathcal{L} = -\sum_i \left[ \frac{(x_i - M_i)^2}{2(e_i^2 + \sigma^2)} + \ln \sqrt{2 \pi (e_i^2 + \sigma^2)} \right],
\end{equation}
where $M_i$ is the predicted velocity (based on either the sinusoidal or constant term models), $x_i$ and $e_i$ are the velocity and error values for each measurement $i$, and $\sigma$ is a systematic uncertainty term added to the measured uncertainties in quadrature. In both cases the fit gave a value for $\sigma$ of around 5\,{\kms}, with the sinusoidal fit also having a semi-amplitude of 5\,{\kms}. We then performed a likelihood ratio test and found that we could not reject the constant fit and hence we have not detected any clear velocity variations in this system.

We placed an upper limit on $K_\mathrm{WD}$ by artificially injecting sinusoidal signals into our data with increasing amplitudes. We then re-ran the same fit and likelihood ratio test (with a fixed value of $\sigma=5$\,\kms) until we could reject the constant model. We found that we should have detected any signal with a semi-amplitude of $\geq 9$\,{\kms} and therefore we set an upper limit for the radial velocity semi-amplitude of the white dwarf in ZTF\,J1230$-$2655 of $K_\mathrm{WD} < 9$\,{\kms}, which is shown as the dashed line in the right-hand panel of Figure~\ref{fig:1230_trail}.

\subsubsection{White dwarf spectral fit}

\begin{figure}
  \begin{center}
    \includegraphics[width=\columnwidth]{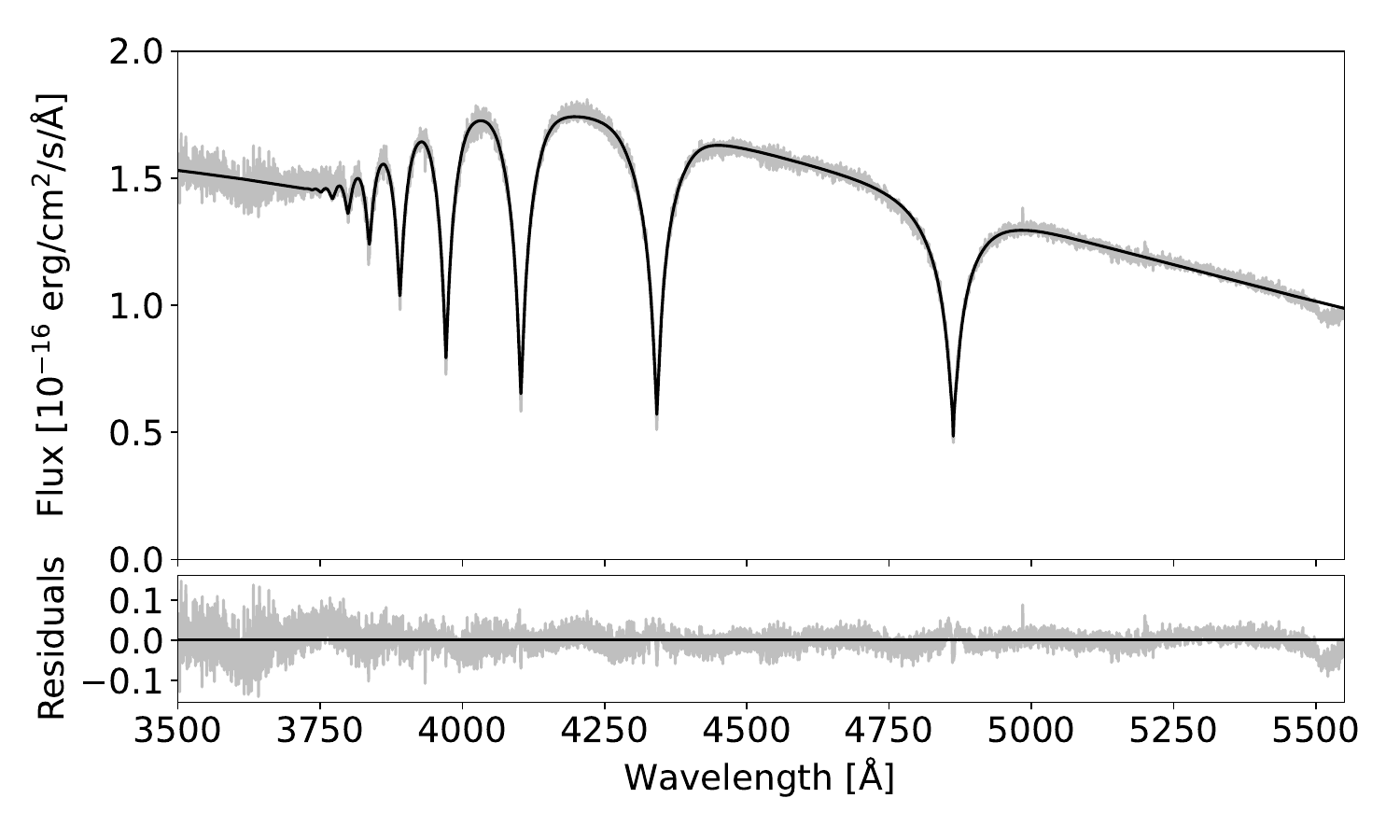}
    \caption{Averaged X-shooter UVB arm spectrum of ZTF\,J1230$-$2655, corrected only for the systemic velocity of the white dwarf (since no radial velocity variations were detected). The best fit \citet{Koester10} model is overplotted in black and the residuals to the fit are shown in the lower panel.}
  \label{fig:1230_specfit}
  \end{center}
\end{figure}

We averaged together all our X-shooter spectra of ZTF\,J1230$-$2655 in order to fit the white dwarf spectrum. Since we did not detect the motion of the white dwarf, we made no correction for this when averaging. We then fitted this spectrum with \citet{Koester10} models using the same method as described for ZTF\,J1828+2308. The average UVB arm spectrum and best fit model are shown in Figure~\ref{fig:1230_specfit}. After applying the 3D corrections of \citet{Tremblay13}, which have a significant effect on the surface gravity, we found that the white dwarf in ZTF\,J1230$-$2655 has $T_\mathrm{eff}=10000\pm110$\,K and $\log{(g)}=8.07\pm0.05$, where again we have applied the systematic uncertainties based on \citet{Hollands24}. This corresponds to a mass of around 0.65\,{\MSUN} based on the mass-radius relationship of \citet{Bedard20}.

As with ZTF\,J1828+2308, we also detect a weak Ca\,{\sc ii} 3934\,{\AA} absorption feature. Although we cannot rule out that this line moves with the white dwarf (since we don't detect the white dwarf's motion), the narrowness of the line leads us to conclude that this is likely an interstellar absorption feature. No other lines except the hydrogen Balmer series are detected in the X-shooter spectrum of ZTF\,J1230$-$2655.

\subsubsection{Light curve fit}

\begin{figure}
  \begin{center}
    \includegraphics[width=\columnwidth]{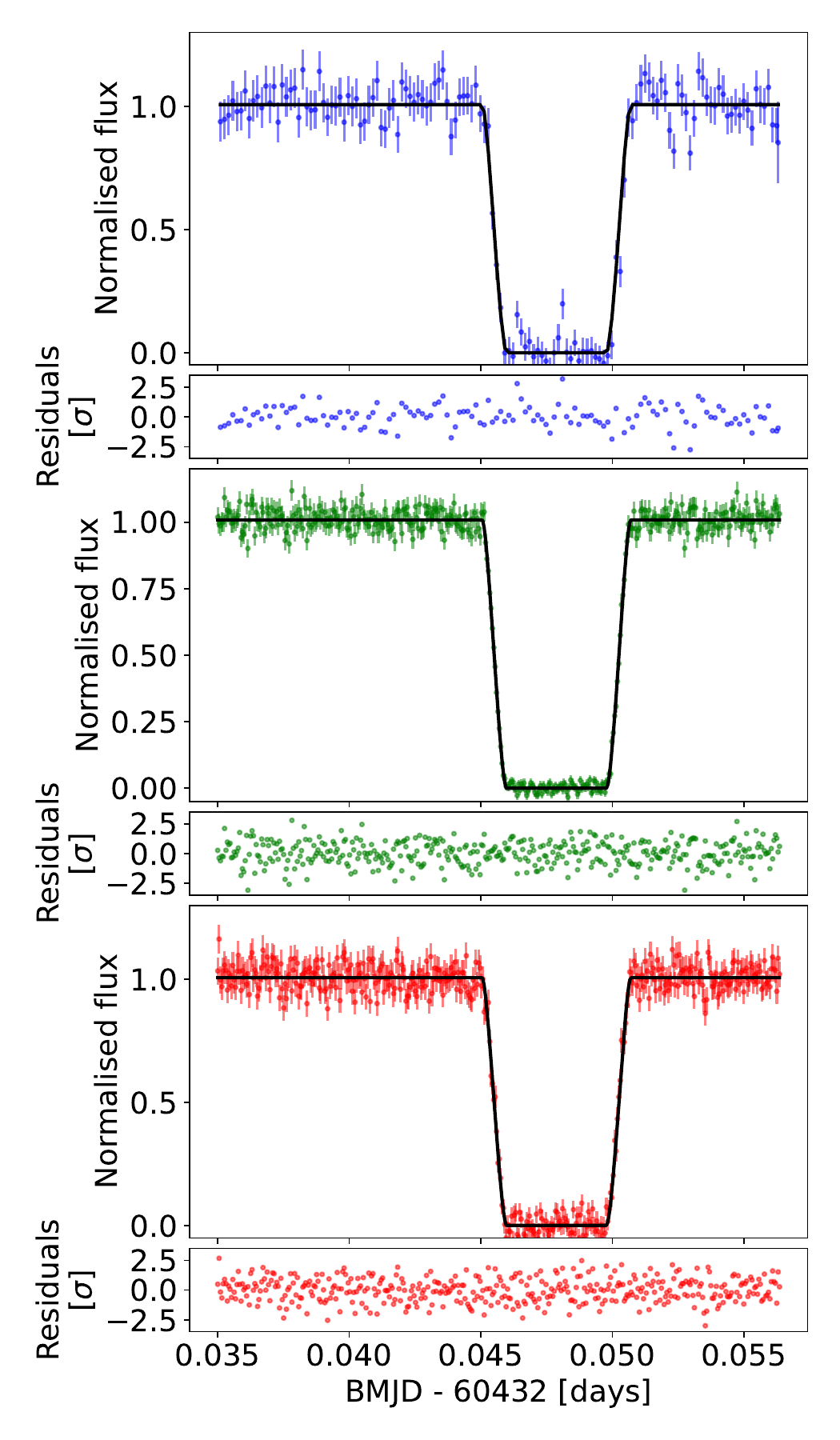}
    \caption{NTT+ULTRACAM eclipse light curves of ZTF\,J1230$-$2655 in the $u_s$ (top), $g_s$ (centre) and $i_s$ (bottom) bands along with the best fit models (black lines). Residuals to the fits are shown beneath each panel (in standard deviations).}
  \label{fig:1230_lcurve}
  \end{center}
\end{figure}

\begin{figure}
  \begin{center}
    \includegraphics[width=\columnwidth]{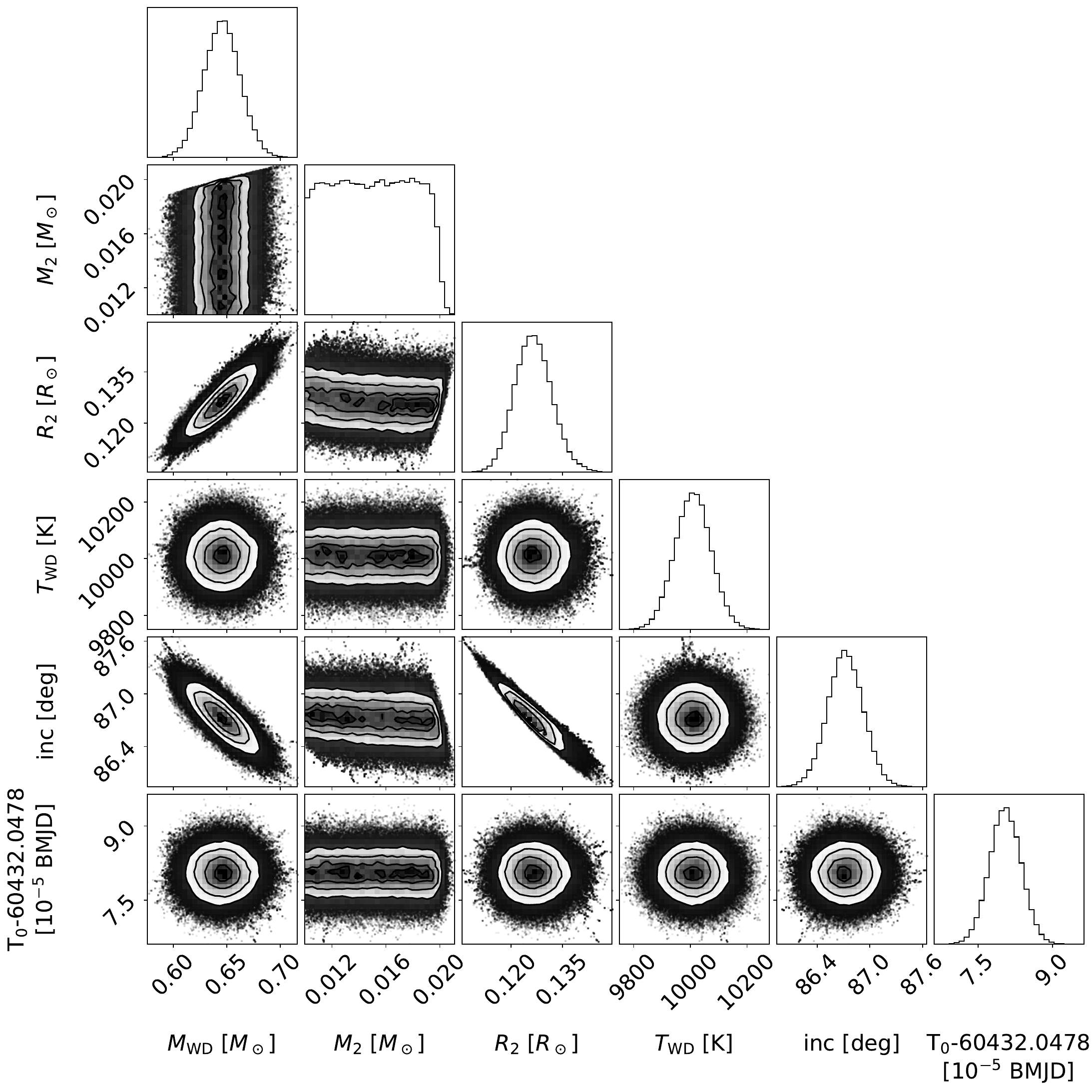}
    \caption{Posterior probability distributions for model parameters obtained through fitting the NTT+ULTRACAM light curves of ZTF\,J1230$-$2655.}
  \label{fig:1230_lcurve_corner}
  \end{center}
\end{figure}

As with ZTF\,J1828+2308, our high-speed multi-band ULTRACAM eclipses of ZTF\,J1230$-$2655 can be used to place strong constraints on the stellar and binary parameters. However, in this case we have no direct measurement of $K_\mathrm{WD}$, just an upper limit, and no constraints on $K_\mathrm{BD}$ at all, since we did not detect any irradiation-driven emission lines in the spectra of ZTF\,J1230$-$2655. Therefore, a full model-independent fit is not possible in this case and we must rely on the use of a theoretical mass-radius relationship for the white dwarf in order to break the degeneracy between the two radii and orbital inclination. Even with the use of a mass-radius relationship, the fact that we only have an upper limit on $K_\mathrm{WD}$ means that we can only place an upper limit on the mass of the companion to the white dwarf.

The fit was performed in the same way as for ZTF\,J1828+2308, using the {\sc lcurve} code, the only differences being that we removed $R_\mathrm{WD}$ as an input and instead computed this value from the mass, temperature and \citet{Bedard20} mass-radius relationship. We used the spectroscopically determined values of the white dwarf effective temperature and surface gravity as Gaussian priors. We also calculated $K_\mathrm{WD}$ for each model using Equation~\ref{eqn:kep3} and placed a uniform prior on this value with an upper limit of 9\,{\kms}, to keep it consistent with our analysis of the radial velocity data. The combination of these constraints meant that we could precisely measure the orbital inclination and radius of the companion, while the white dwarf parameters are determined by the spectroscopic fit, and the mass of the companion is constrained by Equation~\ref{eqn:kep3} and the upper limit on $K_\mathrm{WD}$. We performed an MCMC fit of the light curves in all three bands simultaneously using {\sc emcee} using 100 walkers, with a burn-in period of 2000 and 10000 production steps.

The ULTRACAM eclipse light curves of ZTF\,J1230$-$2655 along with the best fit model are shown in Figure~\ref{fig:1230_lcurve}, the best fit parameters are listed in Table~\ref{tab:params} and the MCMC parameter distributions are shown in Figure~\ref{fig:1230_lcurve_corner}. We find that the companion to the white dwarf in ZTF\,J1230$-$2655 has a mass of $<0.0211$\,{\MSUN} (22.1\,{\MJUP}), putting it in the same mass range as the companion to ZTF\,J1828+2308 and the transiting planet candidate around WD\,1856+534. Indeed, with the present dataset we cannot rule out the possibility that this object has a planetary mass, although for the remainder of this paper we will refer to it as a brown dwarf. Furthermore, just like ZTF\,J1828+2308, the mass of the white dwarf in ZTF\,J1230$-$2655 ($0.646\pm0.017$\,\MSUN) implies that the progenitor of the white dwarf evolved all the way to the AGB before the common envelope phase occurred.

\section{Discussion}

\begin{table}
 \centering
  \caption{Stellar and binary parameters for ZTF\,J1230$-$2655 and ZTF\,J1828+2308. Distances are from \citet{BailerJones21}. For ZTF\,J1828+2308 $K_\mathrm{BD}$ is the corrected value of the centre of mass velocity of the brown dwarf, based on the emission line velocity $K_\mathrm{emis}=343.4\pm1.2$\,{\kms}. The white dwarf in ZTF\,J1230$-$2655 is forced to follow the mass-radius relationship of \citet{Bedard20}. The quoted RLFFs are the volume-average Roche lobe fill factors.}
  \label{tab:params}
  \tabcolsep=0.12cm
  \begin{tabular}{@{}lccc@{}}
  \hline
  Value & Unit & ZTF\,J1230$-$2655 & ZTF\,J1828+2308 \\
  \hline
  RA & hours              & 12:30:16.59 & 18:28:48.77 \\
  Dec & deg               & $-$26:55:51.34 & +23:08:38.39 \\
  Gaia $G$ & mag          & $18.851\pm0.003$ & $17.996\pm0.003$ \\
  Gaia $G_\mathrm{BP}-G_\mathrm{RB}$ &mag & $0.20\pm0.04$ & $-0.11\pm0.02$ \\
  $D$& pc                 & $193\pm9$ & $204\pm5$ \\
  $P_\mathrm{orb}$ & days & 0.235\,977\,66(95) & 0.112\,006\,70(87) \\
  $T_0$ & BMJD(TDB)       & 60432.047\,880\,5(32) & 60431.149\,404\,4(2) \\
  $K_\mathrm{WD}$ & \kms  & $<$9 & $10.5\pm1.1$ \\
  $\gamma_\mathrm{WD}$ & \kms  & $11.7\pm1.5$ & $20.5\pm0.9$ \\
  $K_\mathrm{BD}$ & \kms  & - & $367.0\pm1.2$\\
  $\gamma_\mathrm{BD}$ & \kms  & - & $-12.1\pm0.9$ \\
  $i$ & deg               & $86.7\pm0.3$ & $88.9\pm0.4$ \\
  $a$ & \RSUN             & $1.400\pm0.012$ & $0.838\pm0.002$ \\
  $M_\mathrm{WD}$ & \MSUN & $0.646\pm0.017$ & $0.610\pm0.004$ \\
  $R_\mathrm{WD}$ & \RSUN & $0.0123\pm0.0002$ & $0.0131\pm0.0002$ \\
  WD $\log(g)$ & -        & $8.07\pm0.05$ & $8.05\pm0.05$ \\
  $T_\mathrm{eff,WD}$ & K & $10000\pm110$ & $15900\pm75$ \\
  $\tau_\mathrm{cool,WD}$ & Gyr & $0.726\pm0.050$ & $0.162\pm0.003$ \\
  $M_\mathrm{BD}$ & \MSUN & $<$0.0211 & $0.0186\pm0.0008$ \\
  $R_\mathrm{BD}$ & \RSUN & $0.126\pm0.005$ & $0.102\pm0.002$ \\
  RLFF & - & 0.59 & 0.77 \\
  \hline
\end{tabular}
\end{table}

\subsection{Brown dwarf masses and radii}

\begin{figure*}
  \begin{center}
    \includegraphics[width=\textwidth]{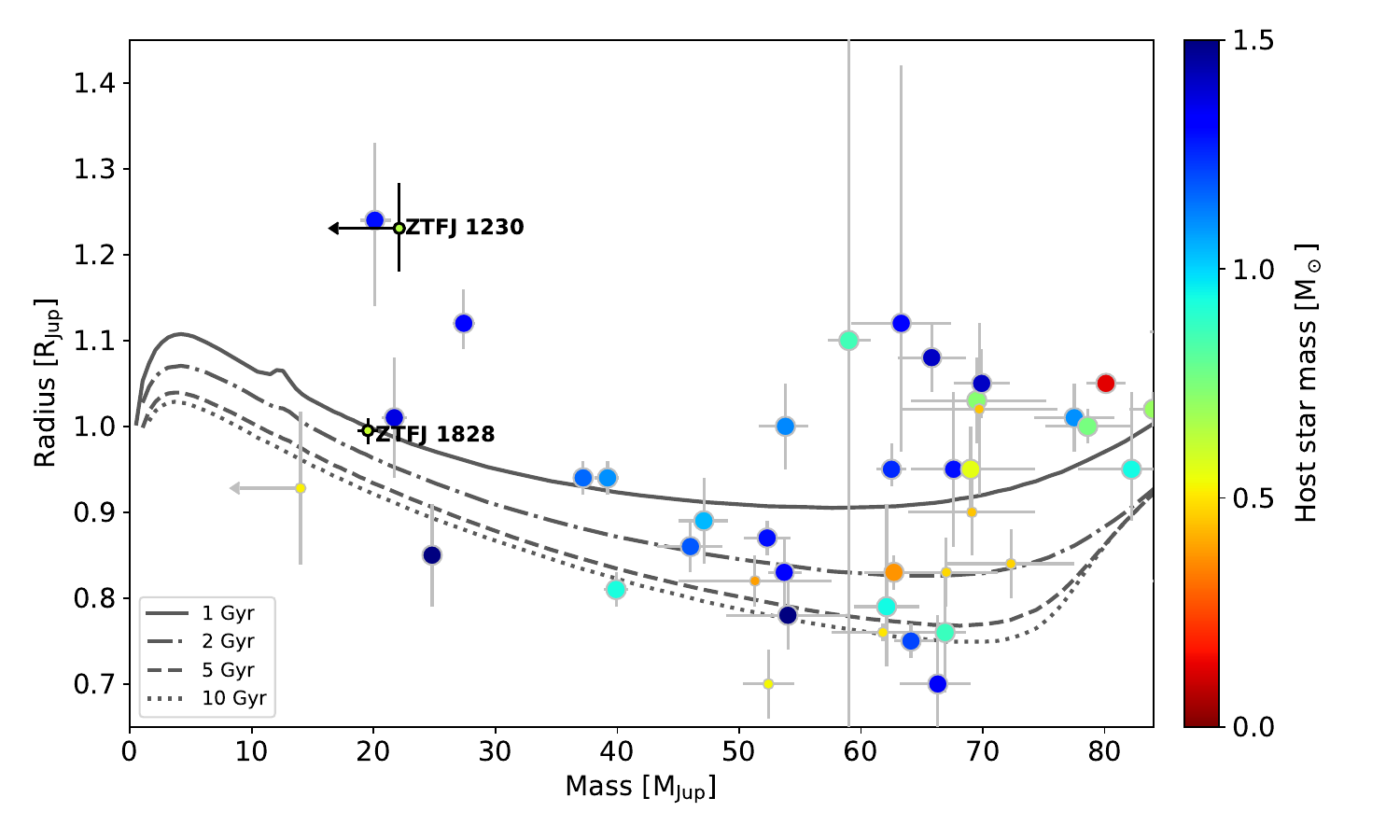}
    \caption{Measured masses and radii of transiting substellar objects with ages $>$1\,Gyr \citep{Littlefair14,Parsons17b,Vanderburg20,Casewell20,vanRoestel21,Grieves21,Psaridi22,Vowell23,Vowell25,Lin23,ElBadry23,Page24,Henderson24,Schaffenroth14,Schaffenroth15,Schaffenroth21}, where the mass of the host star is indicated by the colour of the point. White dwarf and sdB hosts are shown as small points, while main sequence star hosts are shown as large points. ZTF\,J1230$-$2655 and ZTF\,J1828+2308 are highlighted as bold points. Also shown are a number of different evolutionary tracks from \citet{Marley21} for different ages. The masses of the substellar objects in ZTF\,J1230$-$2655 and WD\,1856+534 are upper limits, as indicated by the arrows.}
  \label{fig:BD_radii}
  \end{center}
\end{figure*}

We have measured the mass and radius of the brown dwarf in ZTF\,J1828+2308 independent of any evolutionary models, while the radius and upper mass limit for the brown dwarf in ZTF\,J1230$-$2655 are only dependent on the (well tested) white dwarf mass-radius relationship \citep{Parsons17}. Therefore, these two objects are ideal for testing the mass-radius relationship at substellar masses. Figure~\ref{fig:BD_radii} shows the measured masses and radii of transiting brown dwarfs, where the measurements are independent of evolutionary models. Also shown are a number of evolutionary models for different ages from \citet{Marley21}. Our two systems are highlighted in bold and likely have total ages around 1.5-2\,Gyr (see Section~\ref{sec:evo}). The measured mass and radius of the brown dwarf in ZTF\,J1828+2308 show good agreement with the evolutionary models, with no evidence of over-inflation, despite this object being moderately irradiated. However, despite not having an exact mass value for the brown dwarf in ZTF\,J1230$-$2655, the measured radius is clearly significantly larger than evolutionary models would predict, by around 20 per cent, regardless of the actual mass (if this is a planetary mass object it would still be around 10 per cent oversized). This is despite the fact that this object suffers very little irradiation from the cool white dwarf, and with an orbital period of 5.7 hours, the binary is well detached (only filling 59 per cent of its Roche lobe).

Over-inflated brown dwarfs in close binaries with white dwarfs have been found before. The brown dwarf in the eclipsing system WD\,1032+011 also appears to be over-inflated \citep{Casewell20,French24}, by possibly as much as 30 per cent, given the very old age of the system. The brown dwarfs in the non-eclipsing systems GD\,1400 \citep{Casewell24} and NLTT\,5306 \citep{Amaro23} may also be over-inflated. However, the three other eclipsing white dwarf plus brown dwarf systems with precise mass and radius measurements show good agreement with theoretical models \citep{Littlefair14,Parsons17b,vanRoestel21}. The situation is equally mixed for the population of brown dwarfs around main sequence stars, around half of which appear to be over-inflated. In order to clearly highlight this large variation we used the catalogue of transiting brown dwarfs from \citet{Grieves21} and removed any systems from the that were younger than 1\,Gyr or where there was no good estimate for the age. This means that all the systems plotted in Figure~\ref{fig:BD_radii} should be at or below the 1\,Gyr track in order to have measured radii consistent with theoretical models, but many clearly lie significantly above this track. There is no clear trend with over-inflation and host star mass (hence host star temperature) either, although given that the orbital periods of these systems range from around a day up to more than 100 days the actual irradiation levels vary significantly from object to object.

\begin{figure*}
  \begin{center}
    \includegraphics[width=\textwidth]{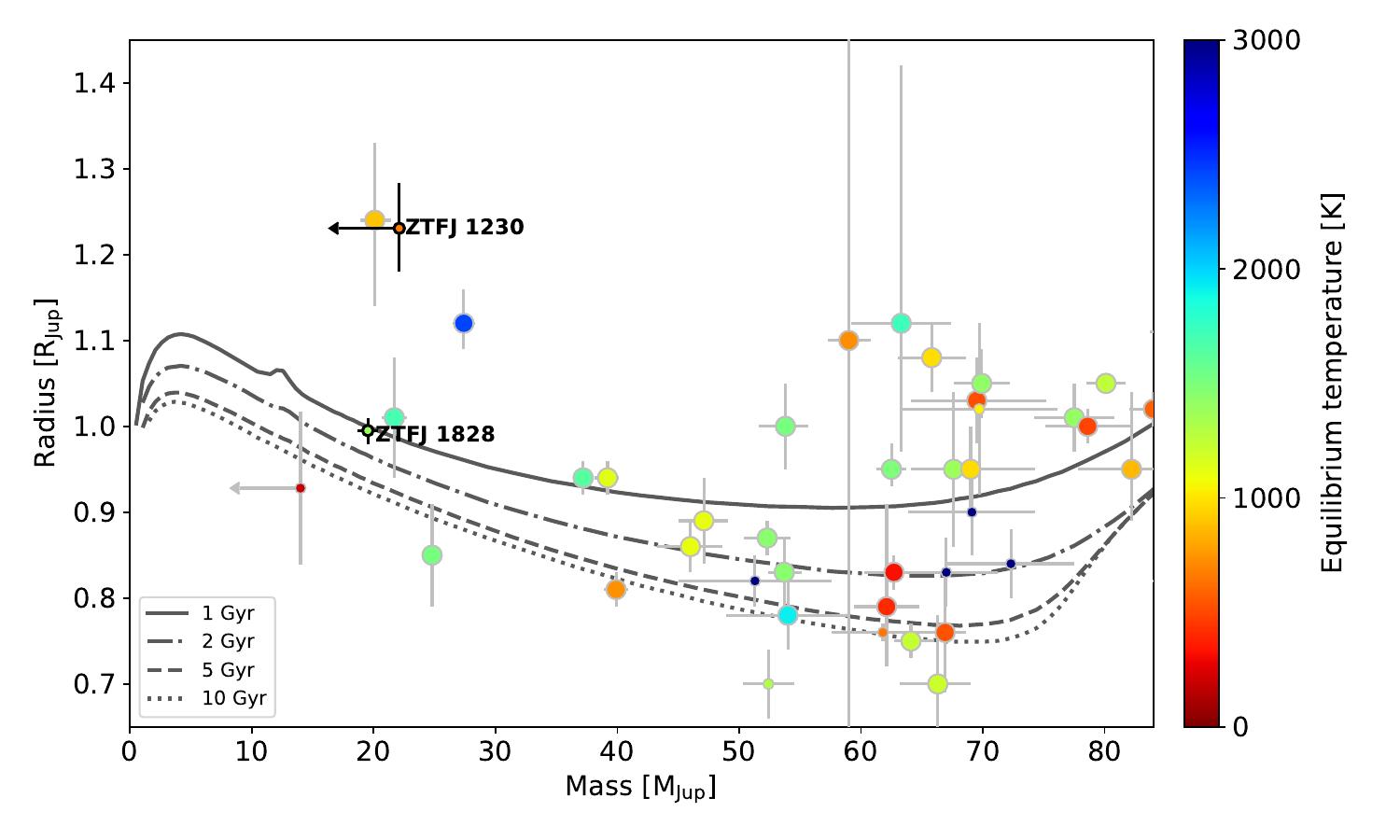}
    \caption{Same as Figure~\ref{fig:BD_radii}, except points are coloured by the equilibrium temperature of the brown dwarf. Note that this temperature ignores the intrinsic luminosity of the brown dwarf, it is purely the equilibrium temperature due to irradiation.}
  \label{fig:BD_radii2}
  \end{center}
\end{figure*}

In order to see if the most over-inflated objects are also the most irradiated, we calculated the time-averaged equilibrium temperature of all of the objects plotted in Figure~\ref{fig:BD_radii} using the relation in \citet{Mendez17}, assuming zero albedo. Note that this neglects the intrinsic luminosity of the brown dwarf, and assumes that it is in thermal equilibrium with the external irradiation. These values are highlighted in Figure~\ref{fig:BD_radii2}, where the highly irradiated objects are clear. However, there is no clear correlation between the irradiation levels and over-inflation, with some highly irradiated objects both oversized (e.g. KELT-1b, \citealt{Siverd12}) and consistent with models (e.g. SDSS\,J1205-0242B, \citealt{Parsons17b}). Likewise there are many objects with low levels of irradiation that are both oversized (e.g. NGTS-19b, \citealt{Acton21}) and consistent with models (e.g. LHS\,6343C, \citealt{Johnson11}). However, the equilibrium temperature only gives a limited view of how these objects are irradiated, since the way the brown dwarf atmosphere responds to the radiation will depend on the wavelength of the radiation. Hotter stars (in particular white dwarfs) emit more of their light at ultraviolet wavelengths, which tends to be absorbed in the higher layers of the brown dwarf's atmosphere potentially leading to temperature inversions \citep{Lothringer20}, while cooler stars emit significant amounts of infrared light, which can pass deeper into the atmosphere before being absorbed, potentially heating (or preventing the cooling of) their interiors \citep{Komacek17}.

Interestingly, the over-inflation of the brown dwarf in ZTF\,J1230$-$2655 is consistent with the trend that, in white dwarf plus brown dwarf binaries, it is the oldest and least irradiated objects that appear to be the most over-inflated \citep{French24}. The over-inflation may therefore be the result of the cumulative effect of irradiation from the white dwarf over a very long time scale slowing down the contraction of these brown dwarfs and the younger systems simply haven't been irradiated for long enough for this effect to be detectable \citep{Casewell20b}. However, the brown dwarf in ZTF\,J1230$-$2655 is so over-inflated that simply delaying the normal contraction is likely not enough to explain its current radius, which is more consistent with a 0.5\,Gyr old brown dwarf (despite a total age of 1.5$-$2\,Gyr, see the next section). Additional mass-radius measurements for close white dwarf brown dwarf binaries should help test this hypothesis, if sufficient numbers (over a wide range of white dwarf cooling ages) can be found and measured.

Alternatively, the brown dwarf in ZTF\,J1230$-$2655 may be over-inflated due to possessing a thick cloud layer or high metallicity \citep{Burrows11}. Unfortunately, since we did not detect any flux from the brown dwarf itself we cannot place any useful limits on the conditions in the atmosphere of this object. Given their low masses, direct detection of the brown dwarfs in either of the systems in this paper will likely be challenging, but might be possible with the James Webb Space Telescope.

\subsection{Evolution} \label{sec:evo}

Given their extremely small separations and circular orbits, both ZTF\,J1230$-$2655 and ZTF\,J1828+2308 are likely the result of CE evolution. There is some debate if the transiting planet candidate around WD\,1856+534 is also a post-CE system or if it migrated into its present orbit after the formation of the white dwarf. Given that WD\,1856+534 is part of a triple system this migration channel seems plausible \citep{Munoz20,OConnor21}, but the CE channel may still be possible \citep{Lagos21}. Indeed, the existence of two additional systems with similarly low mass companions in close orbits around white dwarfs raises the possibility that there does indeed exist a channel for creating such systems via CE evolution. We note that neither of our two new systems appear to form part of a higher-order system at present and there are no common proper motion companions to either system in Gaia DR3 and therefore dynamical scattering or migration seems unlikely in these cases. Indeed, to migrate the brown dwarf in ZTF\,J1828+2308 from several AU (wide enough to avoid the giant star phases) down to its present day separation of just 0.8\,{\RSUN} in only 160\,Myr (the cooling age of the white dwarf) seems extremely challenging. We therefore discuss the evolution of these two systems assuming that they have passed through a CE phase, although as we will see, there are still challenges with this interpretation.

It is also worth noting that the low Roche-lobe filling factors of both systems mean that it is unlikely these are period bounce CVs that have temporarily detached. The short cooling age of the white dwarf in ZTF\,J1828+2308 also supports this, since it would take far longer than the 160\,Myr cooling age for a standard CV to evolve to this phase. While it may be possible to detach period bounce CVs to these kinds of fill factors via the late appearance of a strong white dwarf magnetic field \citep{Schreiber23}, neither of the white dwarfs in these systems have a detectable magnetic field (the lack of any obvious Zeeman splitting in the hydrogen Balmer lines rules out any field stronger than a MG or so), so this mechanism could not have occurred in these cases. The white dwarf mass in both systems is also somewhat lower than is typically found in CVs \citep[e.g.][]{McAllister19,Pala22}. Therefore we consider both systems to be pre-CVs and the current stellar masses are the same as at the end of the CE phase.

\begin{figure*}
  \begin{center}
    \includegraphics[width=0.47\textwidth]{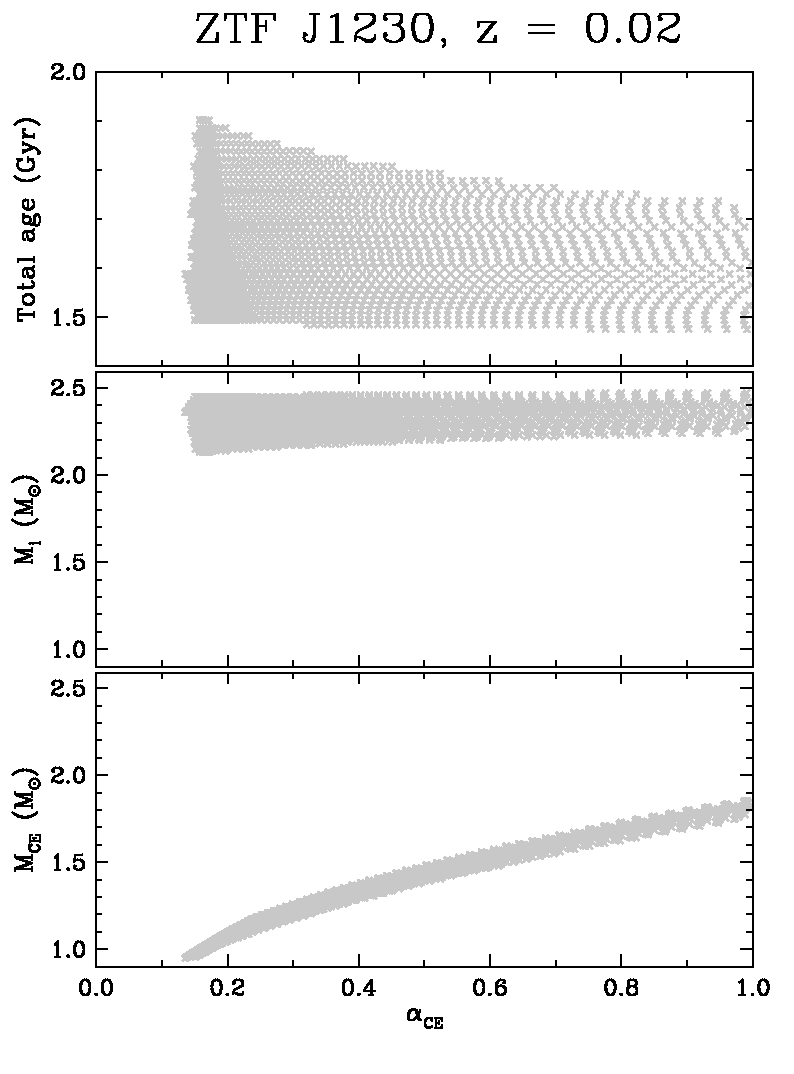}
    \includegraphics[width=0.47\textwidth]{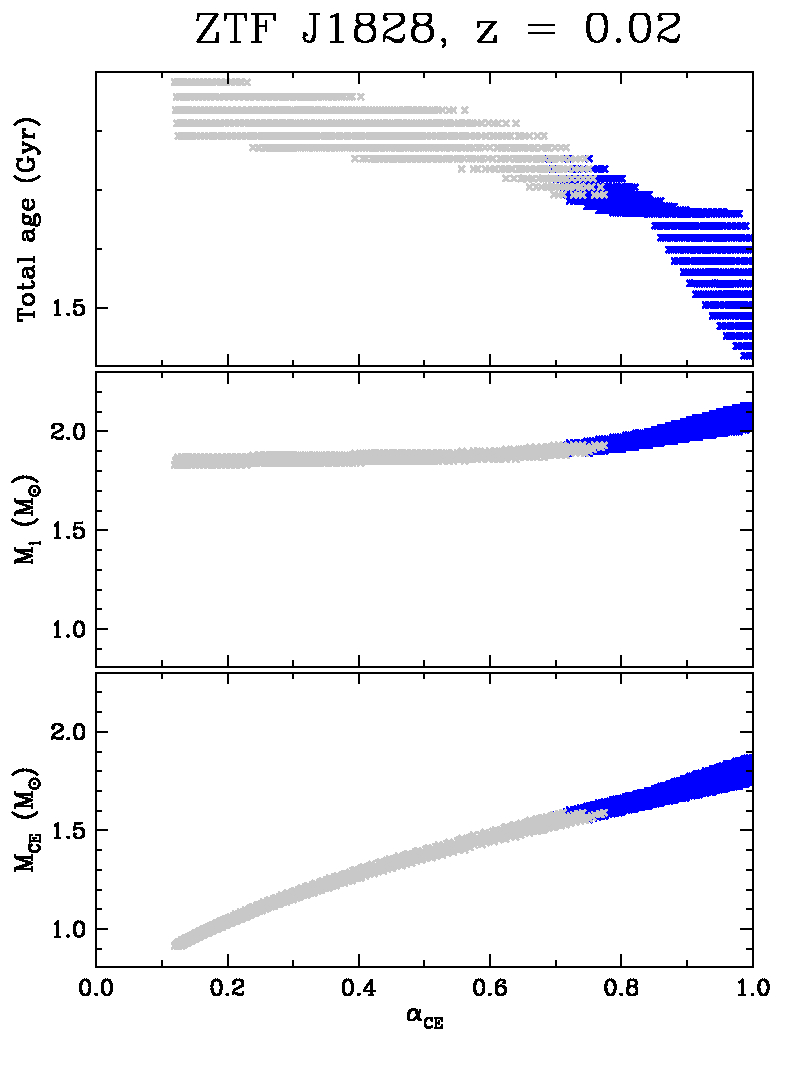}
    \caption{Reconstructions of the evolution of ZTF\,J1230$-$2655 (left, assuming a companion mass of 0.021\,\MSUN) and ZTF\,J1828+2308 (right), assuming a metallicity of $Z=0.02$. Shown are the predicted total age of the system (top), the initial mass of the white dwarf progenitor (middle) and the mass of the white dwarf progenitor at the onset of the common envelope phase (bottom), as a function of the common envelope efficiency parameter, $\alpha_\mathrm{CE}$. Blue points indicate systems where the common envelope occurs during the AGB, before the first thermal pulse, while grey points indicate systems where the common envelope occurs after the first thermal pulse, i.e. during the thermally pulsing (TP)-AGB phase. Only TP-AGB solutions are found for ZTF\,J1230$-$2655.}
  \label{fig:evo}
  \end{center}
\end{figure*}

While the exact details of the CE phase are still poorly understood, it is expected that drag forces acting on the brown dwarf once it has entered the atmosphere of the giant star will cause energy to be transferred from the orbit of the binary into the envelope. If the envelope gains a sufficient amount of energy it will be ejected, leaving behind a much tighter binary. However, this transfer of energy from the orbit to the envelope may not necessarily be very efficient and so only some fraction of the lost orbital energy is actually used to unbind the envelope, typically parametrized as $\alpha_\mathrm{CE}$. A low value of $\alpha_\mathrm{CE}$ implies that only a small fraction of the lost orbital energy is actually used to unbind the envelope, therefore the binary must lose more orbital energy overall in order to eject the envelope. This results in very close binaries emerging from the CE phase or even mergers if the envelope fails to be ejected. To date, studies of both white dwarf plus M dwarf binaries and white dwarf plus brown dwarf binaries have shown that almost all present day systems can be reconstructed with a low value of $\alpha_\mathrm{CE} \simeq 0.2-0.4$ \citep{Zorotovic10,Toonen13,Camacho14,Cojocaru17,Zorotovic22}. Given the low masses (hence low orbital energies) of the brown dwarfs in our systems, they offer a particularly powerful test of whether this low value of $\alpha_\mathrm{CE}$ is universal for these kinds of binaries.

\begin{figure*}
  \begin{center}
    \includegraphics[width=0.47\textwidth]{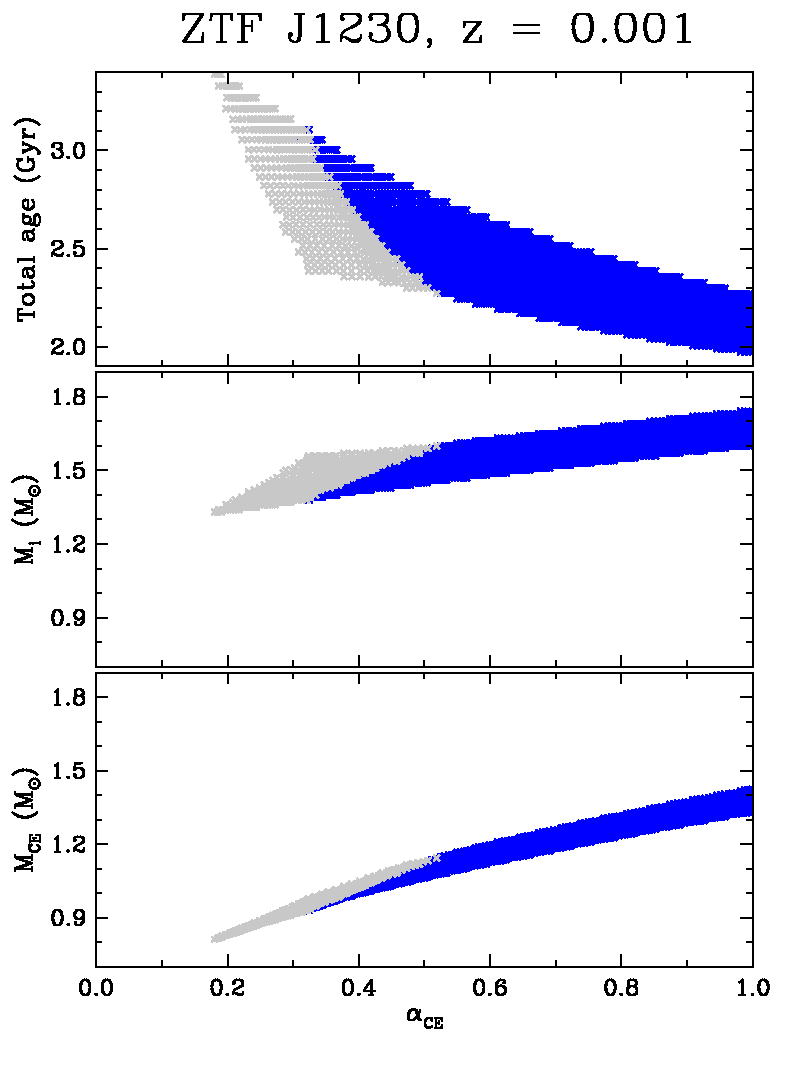}
    \includegraphics[width=0.47\textwidth]{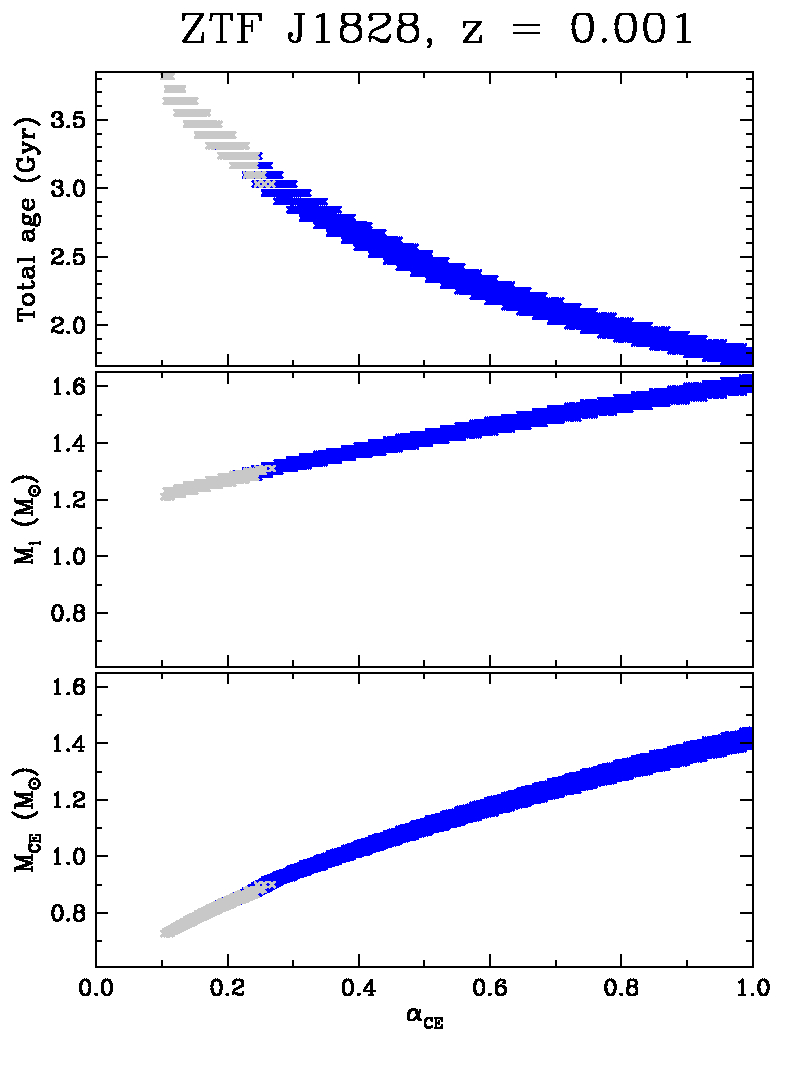}
    \caption{Same as Figure~\ref{fig:evo}, but assuming low metallicity for both systems ($Z=0.001$).}
  \label{fig:evo_lowZ}
  \end{center}
\end{figure*}

We reconstructed the evolution of our systems following the method of \citet{Zorotovic22}, who investigated the CE phase for number of white dwarf plus brown dwarf binaries. In brief, we used the algorithm of \citet{Zorotovic10}, which first uses the measured stellar and binary parameters to correct for any loss of angular momentum (hence reduction in orbital period) since the end of the CE phase. Due to the very low masses of the brown dwarfs it is assumed that only gravitational radiation has reduced the orbital period since the end of the CE phase. The algorithm then searches over a large grid of single star evolutionary tracks to identify any giant stars where the core mass is equal to the measured white dwarf mass. The orbital separation at the onset of the CE phase is then determined using the total mass and radius of the giant star, the mass of the brown dwarf (which we assume does not change) and Roche geometry. The algorithm then uses the energy formalism for CE from \citet{Webbink84}, which relates the energy required to eject the envelope (i.e. its binding energy, $E_\mathrm{bind}$) and the orbital energy released by the binary ($\Delta E_\mathrm{orb}$),
\begin{equation}
E_\mathrm{bind} = \alpha_\mathrm{CE} \Delta E_\mathrm{orb},
\end{equation}
\begin{equation}
E_\mathrm{bind} = - \frac{G M_1 M_{1,e}}{\lambda R_1},
\end{equation}
\begin{equation}
\Delta E_{\text{orb}} = \frac{1}{2} G M_{1,\text{c}} M_2 {\left( \frac{1}{a_i} - \frac{1}{a_f} \right)},
\end{equation}
where $M_1$, $M_{1,c}$ and $M_{1,e}$ are the total, core and envelope mass of the giant star (i.e. $M_{1,c}$ is the mass of the present day white dwarf), $R_1$ is the radius of the giant star, $M_2$ is the mass of the brown dwarf, $a_i$ ($=R_1$) and $a_f$ are the initial and final orbital separations. The structural parameter $\lambda$ is related to the mass distribution of the giant star's envelope and was calculated as in \citet{Claeys14}. This calculation also allows including additional energy sources that reduce the binding energy of the envelope and hence make it easier to expel, e.g. hydrogen recombination energy. However, we did not include these in our reconstructions in order to test whether these low mass systems can still be reconstructed without the need for extra energy sources beyond the orbital energy of the binary, which appears to be the case for other white dwarf plus brown dwarf binaries \citep{Zorotovic22}. $\alpha_\mathrm{CE}$ is treated as a free parameter and we initially fixed the metallicity to the solar value ($Z=0.02$).

The results of this reconstruction are shown in Figure~\ref{fig:evo}. For each model we show the total age of the system (the sum of the pre-CE lifetime and the current cooling age), the initial mass of the white dwarf progenitor and the mass of the progenitor at the onset of the CE phase. For ZTF\,J1230$-$2655 we assumed a mass of the brown dwarf of 0.0211\,{\MSUN}, our measured upper mass limit. In Figure~\ref{fig:evo} we distinguish models where the CE took place on the AGB before the first thermal pulse (blue) and after the first thermal pulse (i.e. on the thermally pulsing AGB - TP-AGB, grey). Given the high white dwarf masses we found no solutions where the white dwarf progenitor was on the first giant branch (FGB). For ZTF\,J1230$-$2655 we only find solutions where the CE occurred during the TP-AGB. Moreover, all of these solutions also require the white dwarf progenitor to have lost a significant fraction of its total mass before the CE phase started. The situation is similar for ZTF\,J1828+2308, although AGB progenitors are possible for very high values of $\alpha_\mathrm{CE}$.

To be consistent with previous constraints of $\alpha_\mathrm{CE}\simeq 0.2-0.4$ would require that both of our systems underwent a CE phase when the white dwarf progenitor was on the TP-AGB and had already lost around a third to half of its original mass. This is perhaps not too surprising a result, given the low masses of the brown dwarfs, since the binding energy of the envelope is already quite low at this phase, allowing even a low mass companion to survive. However, there are potential issues with this interpretation. As pointed out by \citet{Belloni24}, TP-AGB stars generally transfer mass through winds before they fill their Roche lobe, which tends to increase the mass of the companion before the CE. This process is not included in our reconstruction, but given the low masses of the companions in our binaries, they cannot have accreted a substantial amount of mass. Moreover, it has only been 160\,Myr since the CE event in ZTF\,J1828+2308. This is significantly shorter than the thermal timescale of the brown dwarf, so if it had managed to accrete a large amount of material shortly before the CE phase then we would expect it to still be over-inflated, which it is not. This would seem to argue against a CE phase during the TP-AGB for these systems, making it extremely difficult to understand their evolution. However the \citet{Belloni24} study was focused on much more massive companion stars, so it is unclear if their conclusions are applicable to these very low mass objects. For example, the mass accreted from winds should strongly depend on the mass and radius of the accretor. Wind mass loss occurs in all directions, so only the fraction of the wind that can be trapped within the Roche lobe of the accretor will be accreted. Even for the solar-type accretors studied by \citet{Belloni24}, the accreted mass is typically only a few percent of a solar mass. Since brown dwarfs are significantly smaller in both mass and radius, it is unlikely that they can trap a considerable amount of wind material. Assuming a similar scaling rate as Bondi-Hoyle-Lyttleton (i.e. the amount of accreted material scales as $M^2$, \citealt{Edgar04}), would imply that these objects accreted around half a Jupiter mass at most during this phase.

To see if it is possible for a CE to occur before the TP-AGB phase in these binaries we repeated our reconstruction but this time assuming very low metallicity ($Z=0.001$). Low metallicity stars evolve much faster than solar metallicity stars, allowing for less massive progenitors with less envelope mass to expel, hence requiring less orbital energy to be lost from the binary during the CE phase. The results of this are shown in Figure~\ref{fig:evo_lowZ}. In this case there is a much wider range of possible models which undergo the CE phase before the TP-AGB, including some models with low values of $\alpha_\mathrm{CE}$. Therefore, if these are both extremely low metallicity systems then their evolutionary history appears to be broadly consistent with other known white dwarf plus brown dwarf binaries. However, the Gaia kinematics of our systems imply that they are both high probability members of the Galactic thin disk, based on the kinematic criteria of \citet{Bensby03}. Stars in this population tend to have much higher metallicities than $Z=0.001$. Unfortunately, measuring the metallicities of these two systems is not possible at present, but we consider it unlikely that both systems have such low metallicities.

Another possibility for these systems is if there were originally additional objects interior to the brown dwarfs. If these were engulfed shortly before the brown dwarf then they could help facilitate the ejection of the envelope \citep{Bear11,Chamandy21}, allowing lower mass brown dwarfs to survive a CE phase. However, the small number of known main sequence stars with brown dwarfs in the desert (i.e. within a few AU of the star) do not appear to have a significant number of additional objects in closer orbits. This may be due to the fact that the brown dwarf likely formed further out and migrated into the desert, potentially destabilising any nearby objects as it migrated. Therefore, while this is a potential explanation for these systems (as well as a possible evolution for WD\,1856+534; \citealt{Chamandy21}), it would require an architecture unlike any system we have currently detected.

Finally, we note the similarities between our findings and the proposed CE evolution for WD\,1856+534 \citep{Lagos21}. Like our systems, the only solutions found for WD\,1856+534 required the CE to occur on the TP-AGB, although all solutions required a relatively high value of $\alpha_\mathrm{CE} > 0.36$. Moreover, \citet{Lagos21} also required some contribution from additional energy sources to help eject the envelope, which we have not had to include to reconstruct our systems. These differences are partially caused by the lower mass of the object in WD\,1856+534, but are primarily the result of the much longer present day orbital period of WD\,1856+534 (1.4 days) compared to our systems (0.1--0.2 days). Finding more white dwarf plus brown dwarf binaries with longer periods would be particularly useful, to see if WD\,1856+534 is an outlier or consistent with the wider post-CE population.

It may be that these very low mass objects can only survive a CE phase if it occurs right at the end of the evolution of the white dwarf progenitor (i.e. on the TP-AGB), when it has already lost a significant fraction of its original mass and when the envelope is more loosely bound. In this case we would expect the lowest mass brown dwarfs to be only found around CO (or ONe) core white dwarfs. Any similar mass brown dwarf found in a close binary with a He core white dwarf or sdB star would likely pose a significant challenge to reconstruct, since the CE phase would have occurred much earlier, on the FGB. Hydrodynamical simulations place a minimum mass for the companion to eject the envelope at this evolutionary stage of around 30\,{\MJUP} \citep{Kramer20}, well above the brown dwarf masses measured in our systems.

\section{Conclusions}

We have identified two very low mass substellar objects in close, eclipsing orbits around white dwarfs. Using a combination of phase resolved X-shooter spectroscopy and high speed multi-band ULTRACAM and HiPERCAM imaging data we have measured the stellar and binary parameters for these two systems, finding that ZTF\,J1828+2308 contains a 19.5\,{\MJUP} brown dwarf in a 2.7 hour orbit around a hot (15900\,K) CO core (0.61\,{\MSUN}) white dwarf and ZTF\,J1230$-$2655 contains a $<$22.1\,{\MJUP} brown dwarf in a 5.7 hour orbit around a cool (10000\,K) CO core (0.65\,{\MSUN}) white dwarf. We found that the mass and radius of the brown dwarf in ZTF\,J1828+2308 are consistent with evolutionary models, but the brown dwarf in ZTF\,J1230$-$2655 is significantly over-inflated, despite it being the less irradiated of the two. We also reconstructed the past evolution of these binaries, finding that in both cases the common envelope phase likely occurred at the very end of the evolution of the white dwarf progenitor, after the first thermal pulse and when the progenitor had already lost a significant fraction of its original mass. A similar evolutionary history was also proposed for the transiting planet candidate around WD\,1856+534, potentially indicating that common envelope events at this evolutionary stage may allow very low mass brown dwarfs or even planetary mass objects to survive a common envelope phase. Alternatively, both our systems could have very low metallicities, allowing a more standard common envelope phase before the first thermal pulse, or there may have originally been additional bodies interior to the brown dwarf, which were engulfed shortly before the brown dwarf, helping to eject the envelope. Nevertheless, it is clear that post-common envelope binaries with extremely low mass components offer a powerful test of common envelope evolution.

\section*{Acknowledgements}

The results presented in this paper are based on observations collected at the European Southern Observatory under programme IDs 106.D$-$0824 and 113.D$-$0277 and on observations made with the Gran Telescopio Canarias (programme ID GTC119-23B), installed in the Spanish Observatorio del Roque de los Muchachos of the Instituto de Astrof{\'i}sica de Canarias, on the island of La Palma. 

VSD, ULTRACAM and HiPERCAM are funded by the Science and Technology Facilities Council (grant ST/Z000033/1). ARM acknowledges support by the Spanish MINECO grant PID2020-117252GB-I00 and by the AGAUR/Generalitat de Catalunya grant SGR-386/2021. RMO is funded by INTA through grant PRE-OBSERVATORIO. MZ acknowledges financial support from FONDECYT grant number 1221059. IP acknowledges support from a Royal Society University Research Fellowship (URF\textbackslash R1\textbackslash 231496).

%%%%%%%%%%%%%%%%%%%%%%%%%%%%%%%%%%%%%%%%%%%%%%%%%%
\section*{Data Availability}

Raw and reduced X-shooter data are available through the ESO archive. Raw and reduced HiPERCAM data are available through the GTC public archive. Raw and reduced ULTRACAM data will be shared on reasonable request to the corresponding author. 



\bibliographystyle{mnras}
\bibliography{wdbd} % if your bibtex file is called example.bib

%\begin{thebibliography}{99}
%\end{thebibliography}

%%%%%%%%%%%%%%%%%%%%%%%%%%%%%%%%%%%%%%%%%%%%%%%%%%

%%%%%%%%%%%%%%%%% APPENDICES %%%%%%%%%%%%%%%%%%%%%

%\appendix


%%%%%%%%%%%%%%%%%%%%%%%%%%%%%%%%%%%%%%%%%%%%%%%%%%

% Don't change these lines
\bsp	% typesetting comment
\label{lastpage}
\end{document}